\documentclass[a4paper,fleqn,usenatbib,useAMS]{mnras}

\usepackage{newtxtext,newtxmath}

\usepackage[T1]{fontenc}
\usepackage{ae,aecompl}

\usepackage{graphicx}	
\usepackage{amsmath}	
\usepackage{amssymb}	
\usepackage{gensymb}
\usepackage{xfrac}
\usepackage{pdflscape}
\usepackage{xcolor}          

\newcommand{\kms}{\mathrm{km\,s^{-1}}} 
\newcommand{\Msol}{\mathrm{M_\odot}} 
\newcommand{\Msig}{$M-\sigma\,$}
\newcommand{\bfMdot}{$4.0^{+3.5}_{-2.0}\times10^8\,$M$_\odot$}

\title[WISDOM: The SMBH in \mbox{NGC 524}]{WISDOM project -- IV. A molecular gas dynamical measurement of the supermassive black hole mass in \mbox{NGC 524}}

\author[Mark D. Smith et al.]{
Mark D. Smith,$^{1}$\thanks{E-mail: mark.smith@physics.ox.ac.uk}
Martin Bureau,$^{1,2}$
Timothy A. Davis,$^{3}$
Michele Cappellari,$^{1}$
\newauthor{
Lijie Liu,$^{1}$
Eve V. North,$^{3}$
Kyoko Onishi,$^{4}$
Satoru Iguchi$^{5,6}$
and Marc Sarzi$^{7}$}
\\
$^{1}$Sub-department of Astrophysics, Department of Physics, University of Oxford, Denys Wilkinson Building, \\Keble Road, Oxford, OX1 3RH, UK\\
$^{2}$Yonsei Frontier Lab and Department of Astronomy, Yonsei University, 50 Yonsei-ro, Seodaemun-gu, Seoul 03722, \\Republic of Korea\\
$^{3}$School of Physics \& Astronomy, Cardiff University, Queens Buildings, The Parade, Cardiff, CF24 3AA, UK\\
$^{4}$Research Center for Space and Cosmic Evolution, Ehime University, Matsuyama, Ehime, 790-8577, Japan\\
$^{5}$Department of Astronomical Science, SOKENDAI (The Graduate University of Advanced Studies), Mitaka, Tokyo 181-8588, Japan\\
$^{6}$National Astronomical Observatory of Japan, Mitaka, Tokyo, 181-8588, Japan\\
$^{7}$Armagh Observatory and Planetarium, College Hill, Armagh, BT61 9DG, UK
}

\date{Accepted 2019 February 26. Received 2019 February 12; in original form 2018 December 13}

\pubyear{2019}

\begin{document}
\label{firstpage}
\pagerange{\pageref{firstpage}--\pageref{lastpage}}
\maketitle

\begin{abstract}
We present high angular resolution ($0\farcs3$ or $37\,\mathrm{pc}$) Atacama Large Millimeter/sub-millimeter Array (ALMA) observations of the CO(2-1) line emission from a central disc in the early-type galaxy \mbox{NGC 524}. This disc is shown to be dynamically relaxed, exhibiting ordered rotation about a compact $1.3\,\mathrm{mm}$ continuum source, which we identify as emission from an active supermassive black hole (SMBH). There is a hole at the centre of the disc slightly larger than the SMBH sphere of influence. An azimuthal distortion of the observed velocity field is found to be due to either a position angle warp or radial gas flow over the inner $2\farcs5$. By forward-modelling the observations, we obtain an estimate of the SMBH mass of \bfMdot, where the uncertainties are at the $3\sigma$ level. The uncertainties are dominated by the poorly constrained inclination and the stellar mass-to-light ratio of this galaxy, and our measurement is consistent with the established correlation between SMBH mass and stellar velocity dispersion. Our result is roughly half that of the previous stellar dynamical measurement, but is consistent within the uncertainties of both. We also present and apply a new tool for modelling complex molecular gas distributions.

\end{abstract}



\section{INTRODUCTION}
Although representing only a small fraction of its mass, the supermassive black hole (SMBH) now believed to lie at the heart of every galaxy has a major effect on its evolution. Observations over several decades have demonstrated the close relationship between the mass of a SMBH and various properties of its host galaxy \citep[e.g.][]{Magorrian+1998AJ115.2285, Ferrarese+2000ApJL539.9, Gebhardt+2000ApJL539.13, Kormendy+2013ARAA51.511, McConnellMa2013APJ764.184}, implying some form of co-evolution. The nature and explanation of these relationships remain disputed. 

Aside from the aforementioned empirical correlations, the role of SMBHs in galaxy evolution has been explored extensively through theoretical work and simulations \citep[e.g.][]{Silk+2012RAA12.917, Naab+2017ARAA55.59}. Accretion onto a SMBH is considered the most likely explanation for the nuclear activity observed in many galaxies \citep{LyndenBell1969Nature223.690}. Such activity can provide feedback, affecting the evolution of the galaxy. In particular, simulations have shown that the inclusion of feedback from accretion onto a SMBH reproduces a variety of observed galaxy properties \citep[e.g.][]{Meza+2003ApJ590.619, Vogelsberger+2014MNRAS444.1518}, including the stellar mass function \citep[e.g.][]{Croton+2006MNRAS365.11, Schaye+2015MNRAS446.521} and metallicity \citep[e.g.][]{Choi+2017ApJ844.31}. Most recently, observations have found correlations between SMBH masses and galaxy star formation histories, further emphasising the potential role of feedback on a galaxy's stellar mass assembly \citep{Martin-Navarro+2018Nature553.307}.

Most empirical studies infer SMBH masses from correlations with more easily observed quantities; much fewer carry out dynamical measurements based on the kinematics of orbiting material. The latter require high spatial and velocity resolution to disentangle the dynamical contribution of the dominant stellar mass from that of the SMBH in all but the innermost regions, and have historically been done using the velocities of stars, ionised gas or megamasers. Robust measurements are thus far available for only a relatively small number of galaxies \citep[$230$ galaxies are listed in][of which $\approx70$ are upper limits]{vdBosch2016ApJ831.134}.

In the millimetre-Wave Interferometric Survey of Dark Object Masses (WISDOM) project, we are developing a new alternative to the stellar and ionised gas dynamical tracers, through high spatial resolution measurements of the kinematics of molecular gas, available routinely with the Atacama Large Millimeter/Submillimeter Array (ALMA). This method was first demonstrated by \cite{Davis+2013Nature494.328}, in which interferometric observations of the molecular gas disc in \mbox{NGC 4526} allowed the measurement of the SMBH mass via dynamical modelling. Over the past few years, the technique has been characterised \citep{Davis2014MNRAS443.911, Yoon2017MNRAS466.1987} and applied to galaxies across the Hubble sequence, both active and inactive \citep{Onishi+2015ApJ806.39, Barth+2016ApJ823.51, Onishi+2017MNRAS468.4663, Davis+2017MNRAS468.4675, Davis+2018MNRAS473.3818}.

In this paper, we present an estimate of the SMBH mass at the centre of \mbox{NGC 524}, extending our previous techniques to account for its complicated gas distribution, and considering evidence for a kinematic warp in the disc. In Section \ref{sec_target}, we introduce \mbox{NGC 524} and previous studies of it. In Section \ref{sec_observations}, we discuss our observations, their calibration, reduction and imaging, and how they are optimised for our science goals. We then move on to make an estimate of the SMBH mass using our existing techniques in Section \ref{sec_parametricSMBH}, and discuss the systemic uncertainties on our measurement. In Section \ref{sec_SSFit} we modify our techniques by constraining the gas distribution directly using our data, demonstrating and validating a technique which will be needed for future studies of galaxies with complex gas distributions. Section \ref{sec_harmonics} discusses the presence of a non-axisymmetric perturbation in the observed velocity field, and we conclude briefly in Section \ref{sec_conclusion}.

\section{\mbox{NGC 524}}
\label{sec_target}
\mbox{NGC 524} is a nearly face-on early-type galaxy with a core stellar light profile \citep{Faber+1997AJ114.1771}. It has an $I$-band effective radius ($R_\mathrm{e}$) of $51\arcsec$ and a stellar velocity dispersion within $1R_\mathrm{e}$ of $\sigma_\mathrm{e}=220\,\kms$ \citep{Cappellari+2006MNRAS366.1126,Cappellari+2013MNRAS432.1709}. It is classified as a fast rotator, with a specific angular momentum within $1R_\mathrm{e}$ of $\lambda_\mathrm{R_e}=0.28$ \citep{Emsellem+2007MNRAS379.401}. A regular central dust disc with flocculent spiral arms is visible in absorption in \textit{Hubble Space Telescope} (HST) images \citep{Silchenko2000AJ120.741}. Throughout this paper we adopt a distance to \mbox{NGC 524} of $23.3\pm2.3\,\mathrm{Mpc}$, as used in other studies, derived using the surface brightness fluctuation distance of \cite{Tonry+2001ApJ546.681} with the Cepheid zero-point of \mbox{\cite{Freedman+2001ApJ553.47}}. At this distance, $1\arcsec$ corresponds to $113\,\mathrm{pc}$.

\mbox{NGC 524's} molecular gas has previously been observed in the ATLAS$^\mathrm{3D}$ project. \cite{Young+2011MNRAS414.940} observed both the CO(2-1) and CO(1-0) lines with the Institut de Radioastronomie Millim\'etrique (IRAM) 30-m telescope, finding a double-horned profile typical of a rotating disc, and measuring a total molecular gas mass of $M_\mathrm{H_2}=(9\pm1)\times10^7\,\Msol$. In parallel, the disc was spatially-resolved using the Plateau de Bure Interferometer \citep[PdBI;][]{Crocker+2011MNRAS410.1197} in the CO(1-0) line, with a resolution of $2\farcs8\times2\farcs6$ ($320\times290\,\mathrm{pc^2}$). 

\mbox{NGC 524} exhibits nuclear activity, and is revealed as a compact radio source at $5\,\mathrm{GHz}$ by the Very Large Array \citep[VLA;][]{Nyland+2016MNRAS458.2221} and Very Long Baseline Interferometry \citep[VLBI;][]{Filho+2004AandA418.429}. This is indicative of the presence of a SMBH, acting as the central engine of this activity through accretion.  An earlier measurement of the SMBH mass was made by \cite{Krajnovic+2009MNRAS399.1839}, based on stellar kinematics obtained with the adaptive optics-assisted Gemini-North telescope. This galaxy therefore offers an important cross-check between the stellar and molecular gas dynamical techniques. \cite{Krajnovic+2009MNRAS399.1839} concluded that the SMBH mass is $M_\mathrm{BH} = 8.3 ^{+2.7} _{-1.3} \times 10^8\,\Msol$ and the stellar mass-to-light ratio in the $I$-band is $M/L_I = 5.8\pm0.4\,\Msol/\mathrm{L_{\odot,I}}$, having assumed an inclination of $20\degree$ from \cite{Cappellari+2006MNRAS366.1126}. 

The radius of the sphere of influence of the SMBH, within which the SMBH exceeds the stellar contribution to the potential, is defined as $R_\mathrm{SOI}\equiv\frac{GM_\mathrm{BH}}{\sigma^2}$, where $\sigma$ is the stellar velocity dispersion. This is argued to be the relevant length scale for measuring the SMBH mass \citep[e.g.][]{Ferrarese+2005SSRv116.523,Davis2014MNRAS443.911}. Using the \cite{Krajnovic+2009MNRAS399.1839} black hole mass and the aforementioned stellar velocity dispersion, the expected sphere of influence radius is $73\,\mathrm{pc}$ ($0\farcs65$).


\section{Data}
\label{sec_observations}
\subsection{Observations and data reduction}
The data presented here are combined observations of the $^\mathrm{12}$CO(2-1) line in \mbox{NGC 524} from ALMA, using both the $12$-m array and $7$-m Atacama Compact Array (ACA; also known as the Morita array). Data were taken as part of the WISDOM project's observing programmes 2015.1.00466.S (PI: Onishi), 2016.2.00053.S (PI: Liu) and 2017.1.00391.S (PI: North). The $12$-m data span baselines from $15\,\mathrm{m}$ to $1.3\,\mathrm{km}$, providing the high spatial resolution required for our project, and were taken in four tracks on the 26th March 2016, 17th July 2016, 2nd May 2017, and 16th September 2018. The ACA observations, on shorter baselines from $9$ to $48\,\mathrm{m}$ and providing sensitivity to more extended gas structures, were taken in a single track on 25th June 2017. The total on-source time achieved was 2.2 hours with the $12$-m array and 0.3 hours with the ACA. 

For both arrays, a spectral window was positioned to observe the redshifted $J=2$ to $J=1$ transition of $^\mathrm{12}$CO at a velocity resolution of $\approx\,1\mathrm{\,km\,s^{-1}}$ over a bandwidth of $\approx2500\mathrm{\,km\,s^{-1}}$. Three additional spectral windows were positioned to observe the continuum emission, each with a bandwidth of $2\,\mathrm{GHz}$ and a lower spectral resolution.   

The data were calibrated using the standard ALMA pipeline, and combined using the \texttt{Common Astronomy Software Applications} \citep[\texttt{CASA};][]{McMullin+2007ASPC376.127} package. 

\begin{landscape}
\begin{figure} \begin{center}
\centering
	\includegraphics[scale=1.18,trim={0cm 0.4cm 0.6cm 0.6cm}]{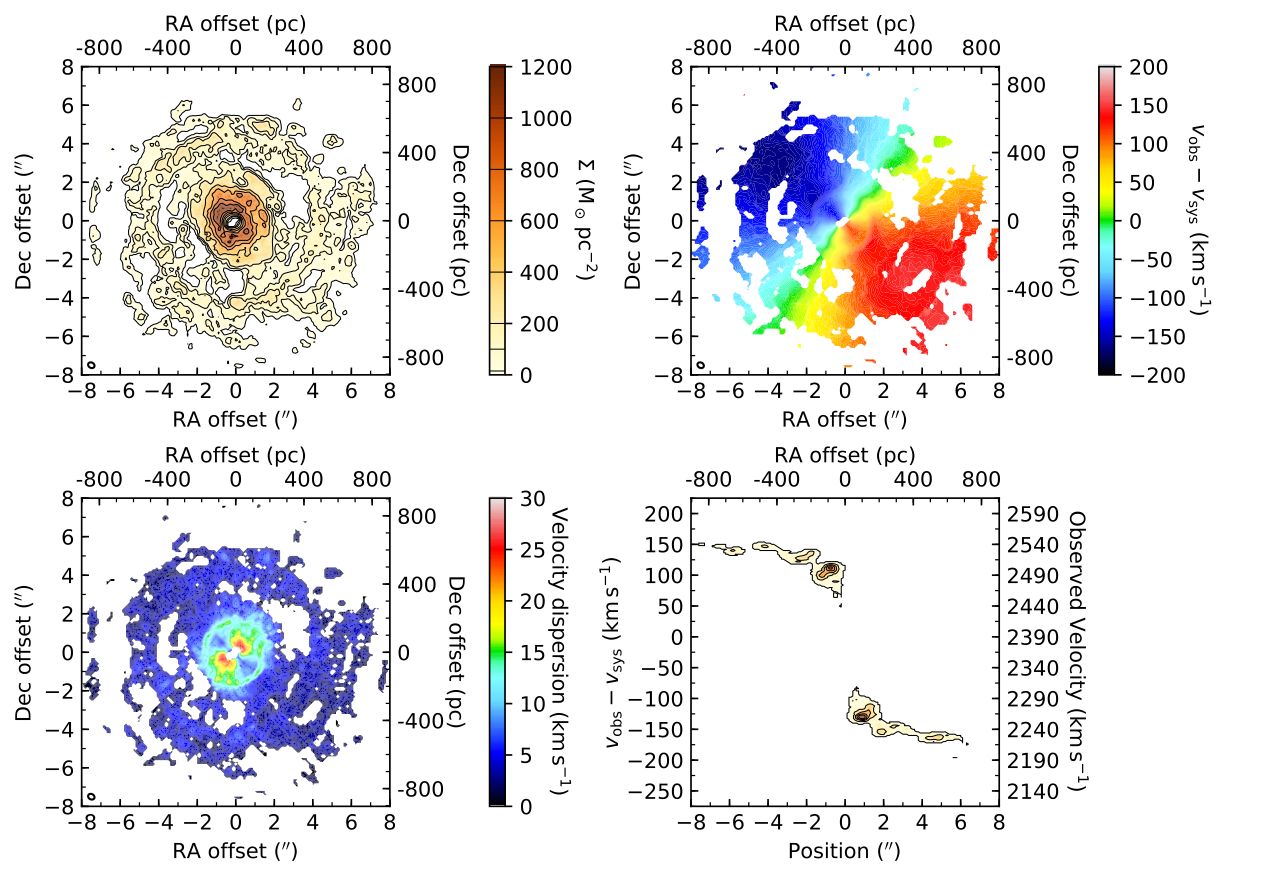}
         \caption{Moment maps of the $^{12}$CO(2-1) emission in \mbox{NGC 524}, from our ALMA data. \textbf{Top-left:} Molecular gas surface density, assuming a CO-to-H$\mathrm{_2}$ conversion factor $\alpha_\mathrm{CO}=4.3\,\Msol\,\mathrm{(K\,\kms)}^{-1}\,\mathrm{pc}^{-2}$. Black contours are from the level at which the noise was clipped, $5\,\mathrm{\Msol\,pc^{-2}}$, and then at $100$, $200$, $400$, $600$, $800$, $1000$, and $1200 \,\mathrm{\Msol\,pc^{-2}}$. \textbf{Top-right:} Mean line-of-sight velocity. \textbf{Bottom-left:} Velocity dispersion. \textbf{Bottom-right:} Kinematic major-axis position-velocity diagram (PVD). In both right panels, $v_\mathrm{obs}$ is the observed line-of-sight velocity and $v_\mathrm{sys}=2390\,\kms$ is the mean systemic velocity.
         We note the presence of a central hole surrounded by a wide ring, from which the gas distribution falls off until a second, outer, ring. We also note the presence of a region of high velocity dispersions at a radius of $\approx2\arcsec$, that correlates with a distortion in the velocity field. There is no evidence of a central Keplerian rise in the PVD, likely because of the central hole.}
    \label{fig_NGC524_maps}
\end{center}
\end{figure}
\end{landscape}

\subsection{Line data}
\label{ssec_linecubes}
To remove the continuum emission from the AGN, a linear fit was made to the line-free channels at both ends of the line spectral window and was subtracted from the \textit{uv} plane using the \texttt{CASA} task \texttt{uvcontsub}. The remaining line data were imaged into an RA-Dec-velocity cube with a channel width of $15\,\kms$. Baselines were weighted by the Briggs scheme with a robust parameter $-1$, weighting towards higher spatial resolution at the expense of sensitivity. The synthesised beam achieved was $0\farcs35\times0\farcs3$ at a position angle of $64\degree$, a factor $\approx7$ improved spatial resolution compared to the \cite{Crocker+2011MNRAS410.1197} observations. This corresponds, at the distance of \mbox{NGC 524}, to a linear scale of $\approx40\times30\,\mathrm{pc}$, so that the predicted radius of the SMBH sphere of influence is resolved with $\approx2$ synthesised beams. The pixel size adopted was $0\farcs1$, such that the beam was approximately Nyquist sampled. The cube size encompasses the array's primary beam spatially, and $\approx15$ channels on either side of the detected line spectrally, but not the entire bandwidth of the spectral window. The sensitivity achieved in the $15\,\mathrm{km\,s}^{-1}$ channels is $0.5\,\mathrm{mJy\,beam^{-1}}$. The cube was cleaned interactively using a manually-defined mask to encompass regions of emission in each channel.

A second cube was created from the same \textit{uv} data with the same parameters but $2\,\mathrm{km\,s}^{-1}$ channels, reaching a sensitivity of $1.1\,\mathrm{mJy\,beam}^{-1}$. This provides the smaller channels required to constrain the gas velocity dispersion, if at the expense of sensitivity (see Section \ref{ssec_gasVelDisp}). The properties of both cubes are tabulated in Table \ref{tab_cubeParameters}.

\begin{table}
	\centering
	\caption{Cube parameters.}
	\label{tab_cubeParameters}
	\begin{tabular}{lcc}
		\hline
		Parameter & $2\,\kms$ cube & $15\,\kms$ cube \\
		\hline
		Image size (px) & $800\times800$ & $800\times800$\\
		Pixel scale ($\arcsec\mathrm{/px}$)& $0.1$ & $0.1$\\
		Channels & 400 & 50\\
		Channel width ($\kms$)& $2$ & $15$\\
		Velocity range ($\kms$) & $2000-2798$&$2000 - 2735$\\
		Synthesised beam ($\arcsec$) & $0.35\times0.3$ & $0.35\times0.3$\\
		Sensitivity ($\mathrm{mJy\,beam^{-1}}$) &1.0 & 0.5\\
		\hline
	\end{tabular}
\end{table}

Moment maps of the integrated flux, mean line-of-sight velocity, and velocity dispersion are shown in Figure \ref{fig_NGC524_maps}. These were made using the smooth-masking method \citep{Dame2011arXiv1101.1499}, whereby the cube is convolved spatially by a Gaussian of the synthesised beam size, is Hanning-smoothed spectrally, and is then clipped at some threshold. This defines a mask that is then applied to the original unsmoothed cube before the moment analysis. 

While the molecular gas appears co-incident with the dust disc seen in optical images, a central hole can clearly be seen. This hole is slightly larger that the predicted sphere of influence of the SMBH, so that we are unlikely to capture the Keplerian rise in the rotation expected as the SMBH begins to exceed the enclosed stellar mass. Similar holes have been seen in smoothed-particle hydrodynamical simulations of the tidal disruption of molecular clouds in galactic nuclei \citep{Trani+2018ApJ864.17}, and in our observations of \mbox{NGC 4429}, but we were still able to obtain a good model of the gas kinematics and thus a good constraint on the SMBH mass \citep{Davis+2018MNRAS473.3818}.

We additionally draw attention to the apparent distortion of the isovelocity contours at a radius of $\approx2\arcsec$ (top-right panel of Figure \ref{fig_NGC524_maps}). The position of this distortion corresponds to that of the ring of increased velocity dispersions in the bottom-left panel of Figure \ref{fig_NGC524_maps}. This feature is considered in detail in Section \ref{sec_harmonics}.

\begin{figure}
	\includegraphics[trim={0cm 1cm 1.5cm 1cm},width=\columnwidth]{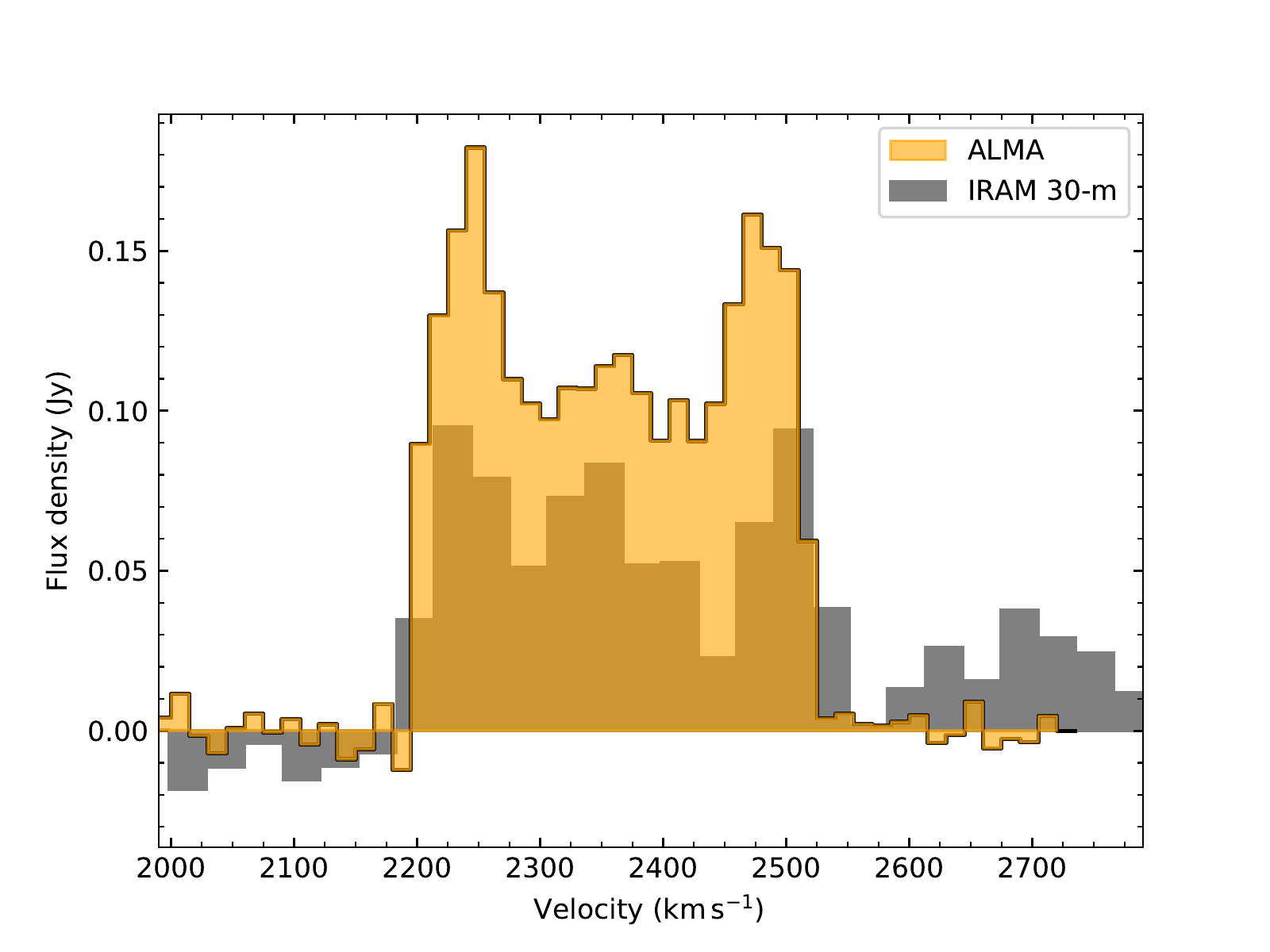}
        \caption{Spatially-integrated spectrum derived from our ALMA observations (orange shading) overlaid on an existing IRAM 30-m telescope spectrum \protect\citep[grey shading;][]{Young+2011MNRAS414.940}. More flux is recovered with ALMA, likely the result of the wider ALMA primary beam, as our observations show that the CO disc extends beyond the 30-m beam. The IRAM 30-m velocities have been converted to the radio convention.}
        \label{fig_NGC524_spec}
        \vspace{-0.3cm}
\end{figure}

Interferometers resolve out flux due to the baselines incompletely sampling the $uv$ plane. We initially try to ascertain how much flux has been recovered by comparing the integrated spectrum derived from our data with that of the IRAM 30-m telescope single-dish observations of \cite{Young+2011MNRAS414.940}, as shown in Figure \ref{fig_NGC524_spec}. These single-dish observations collect all emission within the $10\farcs7$ diameter primary beam. Our integrated flux map (top-left panel of Figure \ref{fig_NGC524_maps}) however shows that the CO disc extends beyond this beam, and Figure \ref{fig_NGC524_spec} indeed shows that ALMA recovers more flux than the IRAM 30-m telescope. While this is encouraging, it is not a proof that all the flux is recovered by ALMA, as we now know that the single-dish spectrum necessarily underestimates the total flux. However, the inclusion of ACA baselines in our data should help to recover any missing flux, and gives a maximum recoverable scale of $29\arcsec$. Since the disc shown in Figure \ref{fig_NGC524_maps} extends only across $\approx16\arcsec$, well within the maximum recoverable scale, and the $uv$ plane is well sampled from this scale to our spatial resolution, we conclude that it is likely that almost all flux has been recovered.

\subsection{Continuum data}
\label{ssec_contIm}
The continuum data, comprising the three other spectral windows and the line-free channels of the high-resolution spectral window (used for the line data), were also imaged. These spectral windows span $17\,$GHz, centred on $237.3\,$GHz, but only sample $7.5\,$GHz of this range due to gaps between the spectral windows. We do not detect continuum emission from the outer parts of the disc, but an unresolved source is detected at the centre of the galaxy, within the hole observed in the line emission. To localise this source, we create a continuum image using the multi-frequency synthesis option of the \texttt{TCLEAN} task in \texttt{CASA}. We use Briggs weighting with a robust parameter of -2, emphasising spatial resolution and achieving a synthesised beam of $0\farcs3\times0\farcs2$ with a sensitivity of  $0.1\,\mathrm{mJy\,beam^{-1}}$. We fit this source with a Gaussian using the \texttt{CASA} task \texttt{imfit}, and it is found to be unresolved, and centred at $01^\rmn{h}24^\rmn{m}47\fs7448$, $+\,9\degree32^\prime20\farcs119$, consistent with the optical centre of the galaxy recorded in the NASA/IPAC Extragalactic Database\footnote{\href{http://ned.ipac.caltech.edu/}{http://ned.ipac.caltech.edu/}} (NED). It has an integrated flux of $8.3\pm0.1\,\mathrm{mJy}$.

Central continuum emission has previously been observed at $3\,\mathrm{mm}$ using the PdBI and was shown to exceed that predicted from dust alone \citep{Crocker+2011MNRAS410.1197}, while at $5\,\mathrm{GHz}$ \cite{Filho+2004AandA418.429} detected a compact radio source on a scale of milli-arcseconds ($0.5\,\mathrm{pc}$ at our adopted distance of $23.3\,\mathrm{Mpc}$). Together these are indicative of continuum emission from an active galactic nucleus (AGN). Our detection is consistent with these positions.

Unlike some of our previous works, we do not present here a spectral energy distribution for this galaxy, since our source is compact. Archival photometric data tabulated in NED show the similarly compact radio observations previously described and \textit{Infrared Astronomical Satellite} (IRAS) observations at $100\,\micron$  and shorter wavelengths. The latter however trace thermal emission from dust on arcminute scales. Although a na\"ive approach to generating an SED shows that our detection appears to be in the Rayleigh-Jeans tail of this dust emission, we do not attempt to infer a dust temperature because of this difference in extent.

\begin{table}
	\centering
	\caption{Parameters of the unresolved continuum source.}
	\label{tab_AGN}
	\begin{tabular}{lc}
		\hline
		Parameter & Value \\
		\hline
		Right ascension & $01^\rmn{h}\,24^\rmn{m}\,47\fs7448\pm0.0001$\\
		Declination & $+9\degree32^\prime20\farcs119\pm0.001$ \\
		Integrated intensity & $8.3\pm0.1$ mJy \\
		Synthesised beam & $0\farcs3\times0\farcs2$\\
		Sensitivity & $0.1\,\mathrm{mJy\,beam^{-1}}$\\
		\hline
	\end{tabular}
\end{table}

\vspace{-0.5cm}
\section{Model fitting}
\label{sec_parametricSMBH}
Our data allow us to measure the SMBH mass in \mbox{NGC 524} by fitting the observed gas kinematics with those derived from a model of the galaxy's mass distribution. This procedure is discussed in detail in \cite{Davis+2017MNRAS468.4675}, so we provide only a brief outline here. 

Within a Markov Chain Monte Carlo (MCMC) framework, we forward-model the observed data cube using the \texttt{Kinematic Molecular Simulation (KinMS)} tool\footnote{\href{https://github.com/TimothyADavis/KinMS}{https://github.com/TimothyADavis/KinMS}} of \cite{Davis+2013MNRAS429.534}, that produces simulated data cubes based on an input gas distribution, circular velocity curve and disc orientation. These simulated data cubes are generated by calculating the line-of-sight projection of the circular velocity for a large number of particles that represent the gas distribution. Additional velocity contributions can be added to account for the velocity dispersion and non-circular motions. The particles are then spatially and spectrally binned, and convolved by the synthesised beam to create a final simulated cube.

We initially assume that the SMBH is located at the position of the unresolved $1.3\,\mathrm{mm}$ continuum source (see Section \ref{ssec_contIm}). The fit is however allowed to vary the exact kinematic centre of the galaxy from this position. We in fact find the centre is very strongly constrained to lie within a pixel of this location, and thus conclude the SMBH is at the kinematic centre of the galaxy's rotation.

The other inputs to our model describe the gas distribution, galaxy potential, and disc geometry. The latter is encapsulated in the overall spatial offsets previously described, a velocity offset, and the disc inclination and position angle relative to the observer.  In the rest of this section, we thus discuss the remaining details of the model (gas distribution and galaxy potential), present the measurements, and estimate the associated uncertainties. 

\subsection{Gas distribution}
In previous works in the WISDOM series, we have assumed some parametric function to describe the gas distribution - typically an exponential thin disc - and have fit the parameters defining this function to the observations as part of the MCMC framework. However, at the resolution achieved the presence of holes, rings and other complex gas morphological features affect our fits, significantly increasing the dimensionality of any model designed to reproduce them.  There are two possible approaches to handle this. In this section, we fit our observations of \mbox{NGC 524} with a coarse axisymmetric parameterisation of the gas distribution, accepting that this will necessarily miss some detailed features of the galaxy. Alternatively, we can directly sample the observed gas distribution to provide input particles to \texttt{KinMS}. We present this new method in Section \ref{sec_SSFit}.

In previous low-resolution observations of \mbox{NGC 524}, an exponential disc was sufficient to adequately describe its gas distribution \citep{Davis+2017MNRAS464.453}. However, our higher angular resolution data reveal a central hole. We therefore adopt an exponential surface density radial profile at large radii, truncated at the edge of a central hole:
\begin{equation}
\label{eq_parametricSB}
I(r) \propto \begin{cases}
         0 & r \leq R_{\text{trunc}}\\
         e^{-\frac{r}{R_0}} & r > R_{\text{trunc}} 
         \end{cases},
\end{equation}
where $R_\text{trunc}$ is the central truncation radius and $R_0$ the scale length of the exponential disc, both of which are parameters in our MCMC fit. We continue to assume the disc is thin and axisymmetric. The central surface brightness of the exponential disc is not used in our model, instead the entire data cube is scaled to an overall integrated intensity, that is also a free parameter within our MCMC code.

\subsection{Stellar mass}
\label{ssec_MGE}
We parametrize the stellar mass distribution of \mbox{NGC 524} using the multi-Gaussian expansion \citep[MGE;][]{Emsellem+1994A&A285.723, Cappellari2002MNRAS333.400} of a HST Wide Field Planetary Camera 2 (WFPC2) F814W image at small radii, and a ground-based $I$-band image from the MDM observatory's 1.3-m McGraw-Hill Telescope at large radii. This MGE model was originally reported in Table B1 of \cite{Cappellari+2006MNRAS366.1126} and is reproduced in Table \ref{tab_MGE} of this work. This is the same model as that adopted in the prior stellar dynamical measurement of the SMBH mass \citep{Krajnovic+2009MNRAS399.1839}. This model was made with no correction for extinction due to the flocculent dust disc in \mbox{NGC 524}, a correction we discuss in Section \ref{ssec_MLGrad}.

We assume the central MGE component, that is unresolved at our angular resolution, to correspond to optical emission from the AGN. It therefore should not contribute additional stellar mass to our model, and we exclude it. We will consider the effect of this choice in Section \ref{ssec_MLGrad}, including this component in the limiting case that all the emission is due to stellar light, and showing that it has a negligible effect on the best-fitting model.

\begin{table}
	\centering
	\caption{Deconvolved 2D MGE components of \mbox{NGC 524}, reproduced from \protect\cite{Cappellari+2006MNRAS366.1126}.}
	\label{tab_MGE}
	\begin{tabular}{ccc} 
		\hline
		$\log_{10}I^\prime_j$ & $\log_{10}{\sigma_j}$ & $q^\prime_j$\\
		(L$_{\odot,I}$pc$^{-2}$) & ($^{\prime\prime}$) & \\
		(1) & (2) & (3) \\
		\hline
		*4.336 & -1.762 & 0.95 \\
		\phantom{*}3.807 & -1.199 & 0.95 \\
		\phantom{*}4.431 & -0.525 & 0.95 \\
		\phantom{*}3.914 & -0.037 & 0.95 \\
		\phantom{*}3.638 & \phantom{-}0.327 & 0.95 \\
		\phantom{*}3.530 & \phantom{-} 0.629 & 0.95 \\
		\phantom{*}3.073 &  \phantom{-}1.082 & 0.95 \\
		\phantom{*}2.450 &  \phantom{-}1.475 & 0.95 \\
		\phantom{*}1.832 &  \phantom{-}1.708 & 0.95 \\
		\phantom{*}1.300 &  \phantom{-}2.132 & 0.95 \\
		\hline
		\end{tabular}
		\parbox[t]{0.45\textwidth}{\textbf{Notes:} The table lists the central surface brightness (column 1), width (column 2), and axial ratio (column 3) of each Gaussian component. The innermost, unresolved Gaussian marked with a star is assumed to relate to emission from the AGN, and is thus omitted from our kinematic fits. Column 1 lists the intensity of each component, Column 2 the width, and Column 3 the axis ratio.}
	
\end{table}

The two-dimensional (2D) MGE parameterisation of the stellar light distribution is used to construct the circular velocity curve of the CO disc. The MGE model can be analytically deprojected to a three-dimensional (3D) distribution given a viewing angle (inclination). In our model, we are able to assume axisymmetry, and the inclination adopted is that of the CO disc.

From this 3D light distribution, a mass distribution is obtained by multiplying by a spatially-constant mass-to-light ratio ($M/L$), another parameter included in our MCMC fit. Use of a constant $M/L$ has been validated by previous kinematic and stellar population studies of \mbox{NGC 524} \citep{Davis+2017MNRAS464.453}, wherein the kinematics of earlier CO(1-0) PdBI observations \citep{Crocker+2011MNRAS410.1197} were fit using the \cite{Krajnovic+2009MNRAS399.1839} SMBH mass, our adopted MGE model and a radially-varying mass-to-light ratio, concluding that an essentially flat dynamical mass-to-light ratio adequately reproduced the disc kinematics. This conclusion was further supported by a uniformly old stellar population.

Given this 3D mass distribution, the potential can be easily calculated by performing the one-dimensional integral given by Equation 12 of \cite{Cappellari2002MNRAS333.400}, and the circular velocity curve directly follows.

\subsection{Bayesian parameter estimation}
\label{ssec_Bayes}
We use the Gibbs sampling MCMC code \texttt{KinMS\_MCMC}\footnote{\href{https://github.com/TimothyADavis/KinMS_MCMC}{https://github.com/TimothyADavis/KinMS\_MCMC}} with adaptive stepping to explore the parameter space. Assuming Gaussian errors, we use the chi-squared statistic as a metric of the goodness-of-fit of the model to the data: 
\begin{equation}
\label{eq_chisq}
\chi^2 \equiv \sum_i\left(\frac{\mathrm{data}_i - \mathrm{model}_i}{\sigma_i}\right)^2 = \frac{1}{\sigma^2}\sum_i(\mathrm{data}_i - \mathrm{model}_i)^2\,\,\,,
\end{equation}
where the sum is performed over all the pixels within the region of the data cube that the model fits, and $\sigma$ is the rms noise as measured in line-free channels of the data cube, that we assume constant for all pixels. Samples are then drawn from the posterior distribution described by the the log-likelihood function $\ln P=-\sfrac{\chi^2}{2}$. Initially, the step size is adaptively scaled to ensure a minimum acceptance fraction is reached and the chain converges, before the entire chain is re-run at a fixed step size to sample the full posterior distribution. The final chain length is $3\times10^5$ steps, with the first $10\%$ discarded as a burn-in phase.

To ensure the chain converges, we always set uniform priors within physical limits (see Table \ref{tab_MCMCresults}, columns 2 and 6), to constrain the range of parameter space the fit can explore. In particular, we draw attention to the prior boundaries on the disc inclination, where the lower bound is the lowest inclination for which the MGE can be analytically de-projected, and to the prior for the SMBH mass, that is uniform in logarithmic space rather than linear space.

\subsubsection{Pixel--to--pixel correlations}
\label{ssec_covar}
As the product of interferometric observations is the source emission convolved by the synthesised beam, the latter oversampled in our data cube, adjacent pixels do not provide independent measures of the $\chi^2$ value. In practice we therefore use a more general form of Equation \ref{eq_chisq}, that includes the inverse of the covariance matrix describing pixel-to-pixel correlations \citep{Barlow1989Wiley}, as discussed in \cite{Davis+2017MNRAS468.4675}. 

The number of elements in this covariance matrix scales as the square of the number of spaxels included, rapidly becoming prohibitively expensive computationally. We therefore only fit the central $6\farcs4\times6\farcs4$ ($720\,\mathrm{pc}\times720\,\mathrm{pc}$) region of our cube. We previously predicted the sphere of influence of the SMBH to be only $0\farcs65$, so this region still provides a very large number of spaxels dominated by the stellar mass distribution, necessary to accurately constrain the stellar mass-to-light ratio.

\subsubsection{Chi--squared uncertainties}
\label{ssec_modifiedBayes}
As our data are noisy, the $\chi^2$ statistic has an additional uncertainty associated with it, following the chi-squared distribution \citep{Andrae2010arXiv1009.2755}. This distribution has a variance of $2(N-P)$, where $N$ is the number of constraints and $P$ the number of inferred parameters. For our data $N$ is very large ($\approx10^5$), so the variance becomes $\approx2N$.

The traditional approach to inferring uncertainties in a single parameter using a $\chi^2$ grid is to select the $1\sigma$ ($68\%$) confidence interval as the contour within which $\Delta\chi^2=1$. However, as \cite{vandenBosch+2009MNRAS398.1117} noted, this approach yields unrealistically small uncertainty estimates due to systematic effects, which can produce variations of $\chi^2$ of the order of its formal error $\sqrt{2N}$. They proposed to increase the $1\sigma$ confidence interval to $\Delta\chi^2=\sqrt{2N}$.

As we are using a Bayesian MCMC approach, rather than $\chi^2$ contours, to achieve the same effect we need to scale the log-likelihood, as done by \cite{Mitzkus+2017MNRAS464.4789}. This is done here by dividing the $\chi^2$ of each model by $\sqrt{2N}$, which is identical to increasing the input errors (the noise in the cube) by $(2N)^{1/4}$, as in \cite{Mitzkus+2017MNRAS464.4789}. This approach appears to yield physically credible formal uncertainties in the inferred parameters, whereas otherwise these uncertainties are unphysically small. Additionally, this remains a conservative estimate for the uncertainty on the SMBH mass, as the mass will be determined by only the innermost pixels, rather than the full $N$.

\subsection{Gas velocity dispersion}
\label{ssec_gasVelDisp}
We include in our model a non-zero velocity dispersion ($\sigma_\mathrm{gas}$), that will become large if the disc is not dynamically cold. We assume that this dispersion is spatially invariant, and only a small linear perturbation on the circular velocity field. Evidence for this is provided by the bottom-left panel of Figure \ref{fig_NGC524_maps}, where in regions unaffected by beam smearing the velocity dispersions are uniformly less than $10\,\kms$, and even in beam-smeared regions is $<30\,\kms$. This is to be compared to the typical (deprojected) rotation velocities ($v_\mathrm{rot}$) exceeding $400\,\kms$.

The main data cube used for our SMBH measurement has channel widths of $15\,\kms$, chosen to improve the signal-to-noise ratio of our data while still adequately probing the gas kinematics. However, as $\sigma_\mathrm{gas}$ is found to be less than a channel width, a better constraint on it can be derived by adopting narrower channels that resolve $\sigma_\mathrm{gas}$ spectrally. We thus opt to perform an initial fit using our cube with $2\,\kms$ channels, finding the best-fitting velocity dispersion, and then fixing this value for subsequent fits to our cube with $15\,\kms$ channels from which we derive the other model parameters. The best-fitting velocity dispersion is then $9.3\,\kms$, fixed in all subsequent MCMC runs.

As suggested above, the low velocity dispersion implies $v_\mathrm{rot}/\sigma_\mathrm{gas}\approx 44$ (assuming $i=20\degree$ and the velocity dispersion to be isotropic), indicating the disc is nearly perfectly rotationally supported. This is consistent with our assumption that the rotation curve is dominated by gravitational forces, and hence traces the galaxy potential.

\subsection{SMBH mass}
\label{ssec_SMBHResult}
As stated above, following the initial fit, we perform a second fit over the 10 remaining parameters describing our model: SMBH mass, $I$-band mass-to-light ratio, gas disc scale length, truncation radius and an overall luminosity scaling, the disc position angle and inclination, and one offset in each of the three cube dimensions (RA, Dec., and velocity). Although most parameters are found to be within the priors, the disc inclination is not, with a best-fitting value of $19\fdg9^{+4\fdg9}_{-0\fdg8}$, where the uncertainties are the $99.7\%$ confidence interval. However, the sample is truncated by the lower bound of the inclination prior. This prior is dictated by the MGE model described in Section \ref{ssec_MGE}, that cannot be deprojected below this inclination. It would be a mistake to ascribe to this minimum inclination a physical significance, since the Gaussian components themselves do not necessarily have physical significance \mbox{\citep{Cappellari2002MNRAS333.400}}. Thus we cannot be confident that the final chain is a true reflection of the posterior, as lower inclinations are not explored.

\begin{figure}
	\includegraphics[width=\columnwidth,trim={0cm 0.5cm 0cm 0cm}]{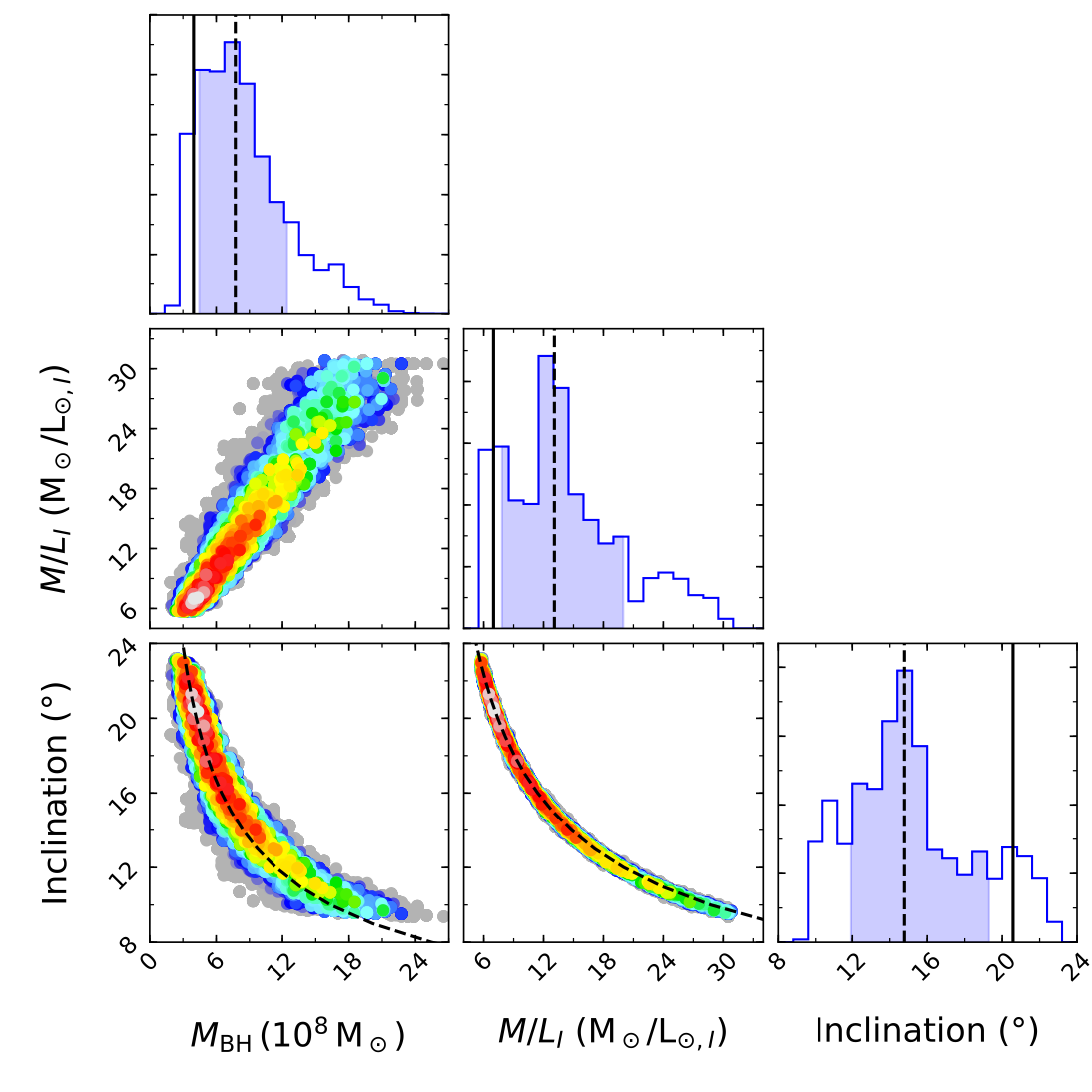}
         \caption{Corner plots showing the covariances between the three key model parameters, from a fit using the circularised MGE model permitting low inclinations. The inclination uncertainty is directly correlated with the uncertainties in both SMBH mass and stellar $M/L_I$. Each point is a realisation of our model, colour-coded to show the relative log-likelihood of that realisation, with white points the most likely and blue least. Grey points are realisations with $\Delta\chi^2>\sqrt{2N}$ relative to the best-fitting model, and are even less likely. Black dashed lines on the scatter plots indicate the predicted effect a varying inclination has on the best-fitting values (following Equation \ref{eq_MbyInc}). The data very closely follows these expected dependencies, showing that the uncertainties are dominated by the inclination. Histograms show 1D marginalisations of each parameter, with black lines denoting the median (dashed) and best-fitting (solid) values. We note that the asymmetry of the posterior means that the most likely value and median are different. The shaded region indicates the $68\%$ confidence interval.}
    \label{fig_NGC0524_incEffect}
\end{figure}

A very small modification to the MGE model allows us to remove this constraint. We circularise the MGE components using the transformation $[I_j^\prime,\sigma_j,q^\prime_j]\rightarrow[I^\prime_j,\sigma_j\sqrt{q^\prime_j},1]$, where $I_j^\prime$, $\sigma_j$, and $q_j^\prime$ are respectively the surface brightness, width, and axis ratio of each Gaussian component, thereby avoiding the need to constrain the inclination prior \citep{Cappellari+2009ApJL704.34}. This transformation effects only a very small change to the MGE model, and ensures that the luminosity and peak surface brightness of each Gaussian is conserved. The priors for, and results of, this fit are shown in columns 2-5 of Table \ref{tab_MCMCresults}. We find a very weak constraint on the inclination, that has an asymmetric posterior and drives a very significant uncertainty in both the SMBH mass and stellar $M/L_I$ derived (see Figure \ref{fig_NGC0524_incEffect}). Indeed, the effect of inclination is so strong that it over-rides the expected inverse correlation between SMBH mass and stellar mass-to-light ratio, that would otherwise conserve the total dynamical mass. This is shown by the positive covariance in the central panel of the left column of Figure \ref{fig_NGC0524_incEffect}. The covariance would be negative at fixed inclination, however as inclination varies the total dynamical mass varies with it, generating the apparent positive covariance in the 2D marginalisation that projects the posterior over inclination.

The effect varying inclination has on SMBH mass measurements can be easily understood by simple arguments. Since we do not directly observe the galaxy's rotation velocity $v_\mathrm{rot}$, but rather the line-of-sight projection $v_\mathrm{rot}\sin i$, the uncertainty associated with inclination will directly affect the black hole mass measurement. We can quantify this effect using a simple circular motion model:

\begin{equation}
     \label{eq_MbyInc}
     M_\mathrm{BH} \propto v^2 \propto \left(\frac{v_\mathrm{obs}}{\sin i}\right)^2.
\end{equation}
As $\sin i$ changes most rapidly at low inclinations it will be a major contributor to the uncertainty on the SMBH mass for the very low inclination disc of \mbox{NGC 524}. Using the circularised MGE model, we find an inclination of $15\degree^{+8\degree}_{-5\degree}$, that from Equation \ref{eq_MbyInc} could change the SMBH mass by $-0.4$ to $+0.3\,\mathrm{dex}$. The median SMBH mass of the accepted models in the MCMC chain is $\log(M_\mathrm{BH}/\Msol)=8.9\pm0.4$, indicating that almost all the uncertainty is due to the inclination uncertainty. The asymmetry of the posterior further means that the best-fit model and the median value of the 1-dimensional (1D) marginalisation are substantially different. The model with the maximum log-likelihood has an inclination of $20.6\degree$ and $\log(M_\mathrm{BH}/\Msol)=8.6$. 

Having said that, we can constrain the inclination using other information. A tilted-ring fit to the velocity field, as described in Section \ref{ssec_harmonicsKinemetry}, yields an average inclination of $21\degree$, with a standard deviation of $6\degree$. Previously, assuming it is intrinsically circular, the dust disc seen in HST images was also found to lie at an inclination of $20\pm5\degree$ \citep{Cappellari+2006MNRAS366.1126}. Since CO is commonly found co-incident with dust \citep[e.g.][]{Crocker+2008MNRAS386.1811,Crocker+2009MNRAS393.1255,Crocker+2011MNRAS410.1197,Young+2008ApJ676.317}, this also provides information on the molecular gas disc inclination \citep{Davis+2011MNRAS414.968}. 

\begin{figure*}
   \centering
   \includegraphics[width=\textwidth]{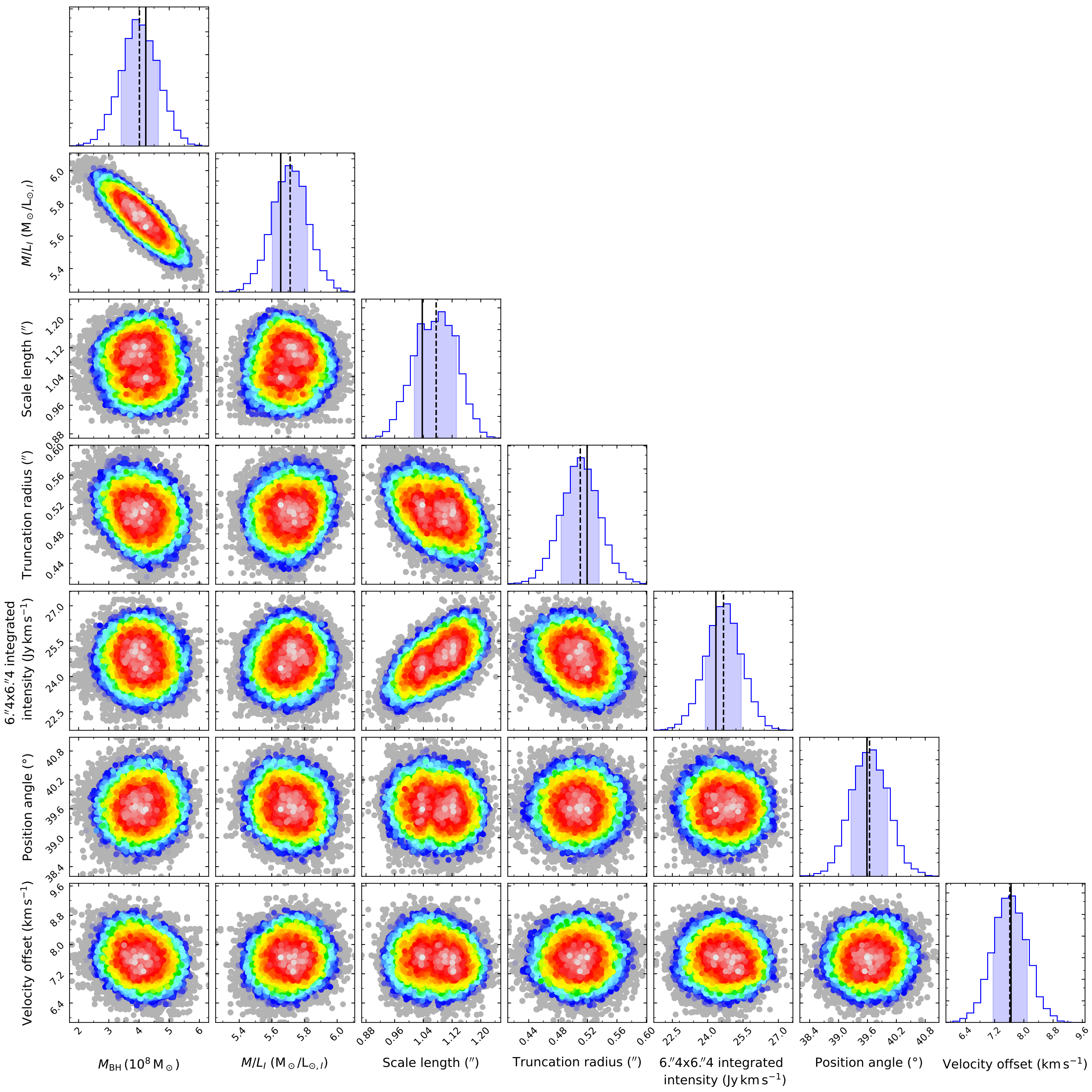}
   \caption{Corner plots showing the covariances between all model parameters, from an MCMC fit at fixed inclination. Each point is a realisation of our model, colour-coded to show the relative log-likelihood of that realisation, with white points the most likely and blue least. Grey points are realisations with $\Delta\chi^2>\sqrt{2N}$ relative to the best-fitting model, and are even less likely. The only significant covariance is between the SMBH mass and the mass-to-light ratio, that corresponds to attributing the dynamical mass across the SMBH and stellar distribution. The weak covariances between integrated intensity, scale length and truncation radius are due to the integrated intensity being the normalisation of the surface brightness profile, and therefore dependent on the exact fits. Histograms show 1D marginalisations of each parameter, with black lines denoting the median (dashed) and best-fitting (solid) values. We note that the asymmetry of the posterior means that the most likely value and median are very slightly different. The shaded region indicates the $68\%$ confidence interval.}
   \label{fig_MCMCcorner}
\end{figure*}

As giving our model extensive freedom to explore inclination does not lead to good constraints when using only our ALMA data, we re-run the fit with a fixed inclination of $20\degree$, adopted from the above arguments. We also take the opportunity to fix the RA and Dec offsets to their previous best-fitting values to reduce the dimensionality of the model. The results of this fit are shown in columns 6-8 of Table \ref{tab_MCMCresults}, and in full corner plots in Figure \ref{fig_MCMCcorner}. With the inclination uncertainty removed, we now obtain much tighter constraints on the SMBH mass and mass-to-light ratio. We then include the inclination uncertainty as a systematic uncertainty in Section \ref{ssec_incUncert}.

Our best-fitting SMBH mass is thus $4.0^{+1.6}_{-1.5}\times10^8\,\Msol$ ($3\sigma$ formal uncertainties) with a reduced $\chi^2$ of $\chi^2_\mathrm{red} = 1.84$. This yields a sphere of influence radius of $36\,\mathrm{pc}$ ($0\farcs31$) that, although marginally spatially resolved by our synthesised beam, is smaller than the hole observed in the CO gas. It is thus unsurprising that we do not see the Keplerian increase in the rotation curve in the very centre of the galaxy. The unusually large uncertainty in the SMBH mass compared to other CO measurements is also likely due to this limitation.

Our best-fitting mass-to-light ratio is $5.7\pm0.3\,\mathrm{\Msol/L_{\odot,I}}$ ($3\sigma$ formal uncertainties). We will consider in Section \ref{ssec_MLGrad} the effect our mass model has on the SMBH mass found.

\begin{figure*}
   \centering
   \includegraphics[trim={1.5cm 0.5cm 2.5cm 1cm},width=\textwidth]{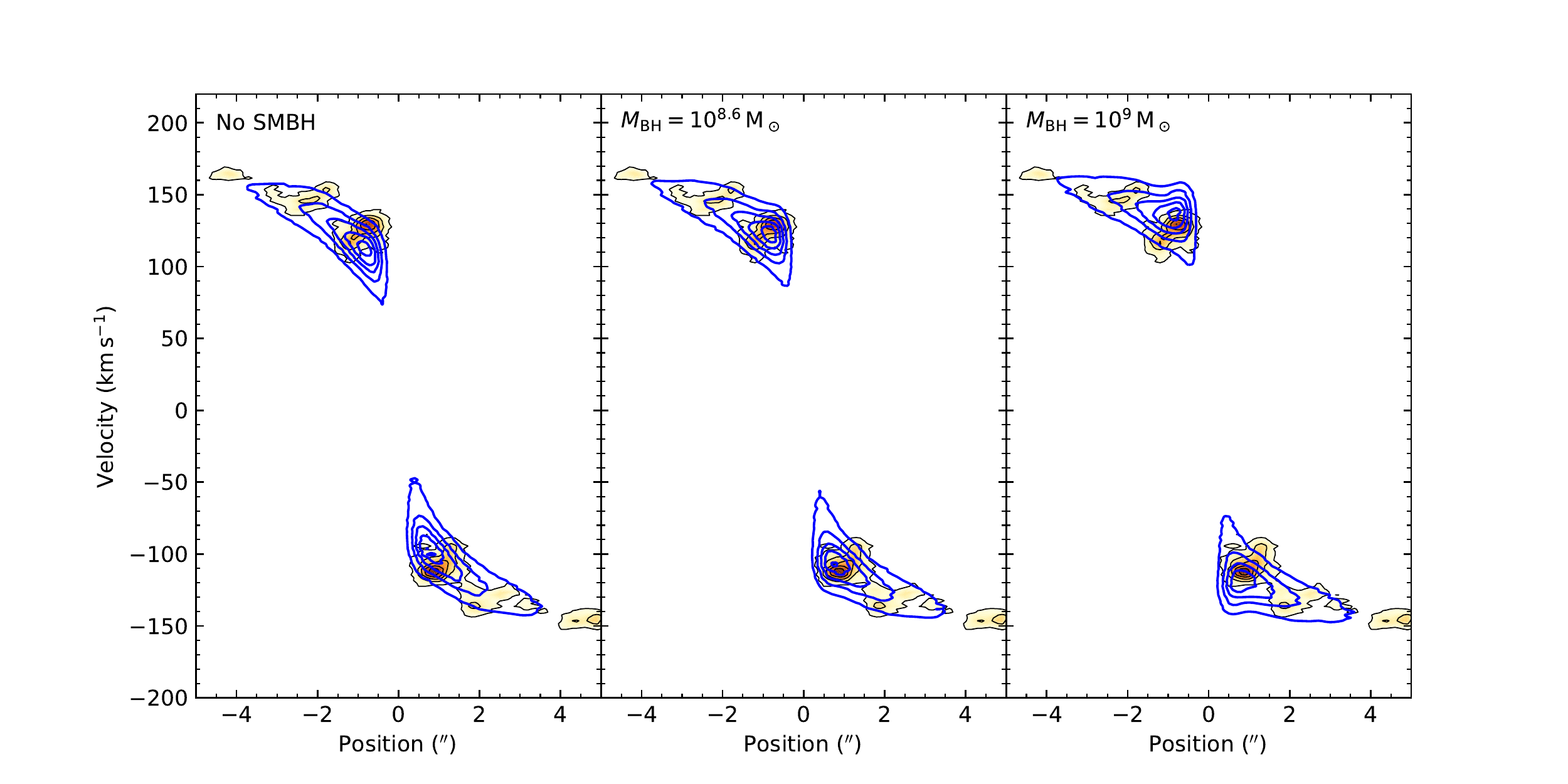}
   \caption{Model position-velocity diagrams along the kinematic major axis of the galaxy (blue contours), showing a model without a SMBH (left), with the best-fitting SMBH (centre) and with an overly large SMBH (right). These are overlaid on the observed PVD previously shown in Figure \ref{fig_NGC524_maps} (orange scales and contours). As can be seen at small radii, the line-of-sight velocities are enhanced compared to a stellar mass-only model, requiring additional central mass to fully account for them.}
   \label{fig_BHPVD}
\end{figure*}

\begin{table*}
	\centering
	\caption{Best-fitting model parameters, with associated formal uncertainties determined using the modified Bayesian sampling approach described in Section \ref{ssec_modifiedBayes}.}
	\label{tab_MCMCresults}
	\begin{tabular}{lcccccccc} 
		\hline
		& \multicolumn{4}{c}{Circularised MGE ($\chi^2_\mathrm{red}=1.84$)} & \multicolumn{3}{c}{Fixed inclination ($\chi^2_\mathrm{red}=1.84$)}\\
		Parameter & Priors & Best fit & Median &$3\sigma$ Error & Priors & Median & $3\sigma$ Error\\
		(1) & (2) & (3) & (4) & (5) & (6) & (7) & (8)\\
		\hline
		\\
		Mass model & & & & & & &\\
		\hline
		$\log$(SMBH mass) ($\Msol$) &  $5\;\rightarrow\;16$ & 8.60 & 8.89 & $\pm0.42$ & $\phantom{0.}5\;\rightarrow\;\phantom{0}12$ & 8.60 & $-0.21,\,+0.15$\\
		Stellar $M/L_I$ ($\mathrm{M_\odot/L_{\odot,I}}$) &  $\phantom{^\star0}1\;\rightarrow\;10^7\,^\star$ &  7.0  & 13.2 & $-7.4,\,+15.0$ & $0.1\;\rightarrow\;\phantom{0}10$ & 5.7 & $\pm0.3$\\
		&&&& & & & \\
		Molecular gas geometry & & & & & & & \\
		\hline
		Scale length ($\arcsec$) & $0.1\;\rightarrow\;\phantom{0}10$ &  1.02 & 1.02 & $\pm0.1$ & $0.1\;\rightarrow\;\phantom{0}10$ & $1.1$ & $\pm0.1$ \\
		Truncation radius ($\arcsec$) & $\phantom{0.}0\;\rightarrow\;\phantom{0}10$ & 0.53 & 0.52 & $\pm0.07$ & $\phantom{0.}0\;\rightarrow\;\phantom{0}10$ & 0.51 & $\pm0.07$\\
                 $6\farcs4\times6\farcs4$ integrated intensity ($\mathrm{Jy\,\kms}$) & $\phantom{0.}1\;\rightarrow\;200$ & 20.8  & 20.8 & $\pm1.5$ & $\phantom{0.}1\;\rightarrow\;200$ &$24.7$ & $-2.0,\,+1.9$ \\
		Gas velocity dispersion ($\kms$) & $\phantom{0.}1\;\rightarrow\;100$ &  9.3  & (fixed) & (fixed) & (fixed) & 9.3 & (fixed)\\
				&&&&&&&\\
		Viewing geometry & & & & & & &\\
		\hline		
		Inclination ($\degree$) & $0.1\;\rightarrow\;\phantom{0}90$ & 20.6  & 14.8 & $-5,\,+8$ & (fixed) & 20 & (fixed)\\
		Position angle ($\degree$) & $\phantom{0.}0\;\rightarrow\;359$ & 39.9  & 39.6 & $\pm1$ & $\phantom{0.}0\;\rightarrow\;359$ & 39.6 & $\pm1$\\
				&&&&&&&\\
		Nuisance Parameters & & & & & & &\\
		\hline
                 Centre RA offset ($\arcsec$) & $\phantom{0}-1\;\rightarrow\;\phantom{00}1$ & -0.12 & -0.12 & $\pm0.04$ &  (fixed) &$-0.12$ & (fixed) \\
                 Centre Dec. offset ($\arcsec$) & $\phantom{0}-1\;\rightarrow\;\phantom{00}1$ & -0.05 & -0.05 & $\pm0.04$ &  (fixed) & $-0.05$ & (fixed)\\
                 Centre velocity offset ($\kms$) & $\phantom{|}-75\;\rightarrow\;\phantom{0}75$ & 7.9 & 7.8 & $\pm1.5$ & $-75\;\rightarrow\;75$ & $7.6$ & $\pm1.2$\\
		\hline
	\end{tabular}
	\parbox[t]{0.95\textwidth}{\textbf{Notes:} The reduced chi--squared value given is that of the model with the best-fitting parameters for each MCMC chain. For the circularised MGE fit, the asymmetric posteriors shown in Figure \ref{fig_NGC0524_incEffect} mean that the minimum chi-squared and the median of the 1D marginalisation of each parameter are not the same, so both are listed. In both fits, the gas velocity dispersion was fixed to the value found in an identical fit to our $2\,\kms$ cube, using the priors listed in column 2. The prior for the mass-to-light ratio marked with a $\star$ is uniform in logarithmic-space for the free inclination fit, where it covers several orders of magnitude, but it is uniform in linear space for the fixed-inclination fit. This avoids unduly favouring high values.}
\end{table*}

\begin{table}
    \caption{Uncertainties estimated by bootstrapping fits to subsets of the data.}
    \label{tab_bootstrap}
    \begin{tabular}{lcc}
    \hline
    Parameter & Best Fit & Uncertainty\\
    &&($3\sigma$)\\
    \hline
    \\
    Mass model & & \\
    \hline
		$\log$(SMBH mass) ($\Msol$) & $8.61$ & $\pm0.21$ \\
		Stellar $M/L_I$ ($\mathrm{M_\odot/L_{\odot,I}}$) & $5.73$ & $\pm0.33$ \\
		&& \\
		Molecular gas geometry & &  \\
		\hline
		Scale length ($\arcsec$) & $1.16$ & $\pm0.12$\\
		Truncation radius ($\arcsec$) & $0.54$ & $\pm0.12$\\
		$6\farcs4\times6\farcs4$ integrated intensity ($\mathrm{Jy\,\kms}$) & $25.1$ & $\pm6.2$\\
		Gas velocity dispersion ($\kms$) & $9.3$ & (fixed)\\
		&&\\
		Viewing geometry & &\\
		\hline
		Inclination ($\degree$) & $20$ & (fixed)\\
		Position angle ($\degree$) & $39.6$ & (fixed)\\
		&&\\
		Nuisance parameters & & \\
		\hline
                Centre RA offset ($\arcsec$) & $-0.12$ & (fixed)\\
                Centre Dec. offset ($\arcsec$) & $-0.05$ & (fixed)\\
                Centre velocity offset ($\kms$) & $7.7$ & (fixed)\\
    \hline
    \end{tabular}
    \parbox[t]{0.45\textwidth}{\textbf{Notes:} The best fitting value listed is the mean of the best-fitting values for each subset, while the uncertainty is 3 times the standard deviation of these values.}
\end{table}

\subsection{Fit uncertainties}
In subsection \ref{ssec_modifiedBayes}, we defined the fit confidence interval used in this work, whereby we modified the standard $\Delta\chi^2$ interval to include the uncertainty in the $\chi^2$ value. As a sanity check of this definition, we make a second estimate of the uncertainties by bootstrapping. We select 16 sub-samples from our cube, constituted of half-cubes bounded by planes at fixed position angles. All planes are at regular angular intervals and pass through the previously established position of the SMBH. For each sample, we minimise the $\chi^2$ defined in Equation \ref{eq_chisq} using \texttt{MPFIT} \citep{Markwardt2009ASPCS411.251}.

For each parameter, we define the overall best fit as the mean and the $1\sigma$ bootstrapped uncertainty as the standard deviation of all the best-fitting values across the 16 sub-samples. The uncertainties thus obtained are listed in Table \ref{tab_bootstrap}. The best-fitting values are consistent with those estimated by the MCMC procedure, and the uncertainty estimates are reassuringly very similar. We are therefore confident that the uncertainties derived using our modified Bayesian approach described in Section \ref{ssec_modifiedBayes} (and listed in Table \ref{tab_MCMCresults}) are reliable estimates of the true uncertainties, and we henceforth adopt them.

\subsection{Systematic effects}
\label{sec_sysErrors}
Our SMBH mass estimate relies on our model being an appropriate model of the data. We thus consider in this section how robust our estimate is against a number of effects: 

\subsubsection{Distance}
Dynamical SMBH mass estimates are systematically affected by the assumed distance ($D$) to the galaxy, with $M_\mathrm{BH}\propto D$. Here we have adopted a distance of $23.3\,\mathrm{Mpc}$, from the surface brightness fluctuation work of \cite{Tonry+2001ApJ546.681}, updated for the Cepheid photometric zero-point of \mbox{\cite{Freedman+2001ApJ553.47}}. The formal uncertainty in this measurement is $\approx10\%$, but since this is a simple normalisation, we follow standard practice and do not include it in our systematic uncertainty.

\subsubsection{Inclination}
\label{ssec_incUncert}
We include in the SMBH mass uncertainty a contribution from inclination uncertainty, by adding the additional contribution predicted from Equation \ref{eq_MbyInc}. For this purpose, we adopt as the inclination uncertainty at $20\degree$ the representative value of $\pm5\degree$ quoted by \cite{Cappellari+2006MNRAS366.1126}. This is slightly more conservative than the uncertainty given by our tilted-ring model, but broadly consistent with the upper bound of the confidence interval from the MCMC fit without the circularised MGE where inclination was allowed to vary. At $25\degree$ inclination, Equation \ref{eq_MbyInc} implies a decrease in $\log (M_\mathrm{BH}/\Msol)$ of $-0.18$ dex, and at $15\degree$ an increase of $+0.24$ dex.

\subsubsection{Mass model}
\label{ssec_MLGrad}
Since the molecular gas disc has a central hole, we do not capture the Keplerian increase of the rotation velocities where the SMBH dominates the mass distribution. We therefore rely on the accuracy of our stellar mass model to constrain the SMBH mass, observing an enhancement of the velocities compared to those expected from the stellar mass alone. 

The $I$-band image used by \cite{Cappellari+2006MNRAS366.1126} to construct the MGE stellar light model discussed in Section \ref{ssec_MGE} may be contaminated by dust extinction, although we expect the extinction in this band to be minimal. It is possible that this will only reduce the total flux without affecting the light distribution, in which case the derived $M/L$ will simply be overestimated, and the SMBH mass derived will be unaffected. However, the flocculent dust distribution visible in \textit{HST} images suggests that the extinction will be irregular. Although correcting for extinction can be challenging, we can make a first-order correction using the prescription of \cite{Cappellari+2002ApJ578.787}. Using the \textit{HST} Planetary Camera F555W and F814W images, we calculate the (V-I) colour for each pixel. We assume the galaxy has an intrinsic (V-I) colour that is a power-law in radius, fit from the dust-free region at the centre of the galaxy, where our CO map exhibits a hole. We then use a standard galactic extinction law \citep{BinneyMerrifield1998book} to correct the $I$-band image for the colour excess in each pixel. This excess is most significant over the region from $0\farcs5$ to $4\arcsec$, diminishing at larger radii. We construct a new MGE model from this extinction-corrected image, and use this model as before to fit the ALMA data cube. Since we are only interested in whether the SMBH mass derived is changed, rather than a full MCMC chain, we only perform a chi-squared minimisation using \texttt{LMFIT}\footnote{\href{https://zenodo.org/record/1699739\#.XF0zvFz7RhE}{DOI:10.5281/zenodo.1699739}}. We find that the SMBH mass is unchanged within our uncertainties, and therefore dust extinction is not a significant effect.

Since our mass model is based entirely on the $I$-band light emitted by the galaxy, we neglect any possible contribution to the potential by non-luminous matter. If this were distributed identically to the stellar mass, the only impact would be that an overestimated stellar $M/L_I$, with no effect on the SMBH mass or other model parameters. However, if centrally concentrated, some of the excess velocities would be attributed to the SMBH, leading us to systematically overestimate the SMBH mass.

Fortunately, on these small spatial scales we expect the stellar mass to dominate the potential. Our observations show that the molecular gas disc extends only to a radius $R\approx8\arcsec$, whereas the $I$-band effective radius $R_{\mathrm{e,}I}=51\arcsec$ \citep{Cappellari+2006MNRAS366.1126}, and we can assume a negligible dark matter contribution over so small a fraction of the galaxy's volume.

Another possible source of mass not considered previously is the interstellar medium (ISM). Interferometric observations of \mbox{NGC 524} by \cite{Osterloo+2010MNRAS409.500} do not detect H{\small{I}} associated with the molecular gas disc, implying that the cold ISM is dominated by molecules. \cite{Young+2011MNRAS414.940} observed \mbox{NGC 524} with the IRAM 30-m telescope, in both CO(1-0) and CO(2-1), reporting a total molecular gas mass of $9\times10^7\,\Msol$ assuming a standard CO-to-H$_2$ conversion factor. However, this mass is spread across the entire disc. We can test if it is significant by comparing it with the total stellar mass present within the same volume, as estimated from our best-fitting model. This is calculated from the integral of the MGE model to the edge of the disc ($R=8\arcsec$) and our best-fitting mass-to-light ratio, yielding a stellar mass of $6\times10^{10}\,\Msol$, thus dwarfing the cold ISM contribution. Additional evidence that we can neglect the molecular gas mass is provided by the presence of the central hole, indicating that the $\mathrm{H_2}$ content around the SMBH is very low. We therefore conclude that the SMBH mass is not biased by the cold ISM at small radii.

In addition, we recall (from Section \ref{ssec_MGE}) that we \textit{a priori} excluded the central MGE component (marked with a * in Table \ref{tab_MGE}), arguing that it is likely due to emission from the AGN, and so should not contribute mass to our stellar distribution. It is however possible that the light in this component is in fact due to stars, that would thus contribute mass to our model, indistinguishable from that of the SMBH as our observations do not resolve them. To test this, we re-fit our data cube fixing all parameters except the SMBH mass to their previous best-fitting values, and including this MGE component. We find no significant change in the best-fitting SMBH mass. This is unsurprising since, even integrated to the edge of the disc and using our best-fitting mass-to-light ratio, this MGE component contributes only $\approx10^6\,\Msol$, or less than $1$ per cent of the previously derived SMBH mass. This potential contribution is thus again dwarfed by both the SMBH mass and the stellar mass due to all the resolved MGE components.

Finally, we assumed that a constant mass-to-light ratio is appropriate across the entire disc. In the outer regions, this has been shown to be a good assumption by \mbox{\cite{Davis+2017MNRAS464.453}}, using the earlier lower-resolution observations of \cite{Crocker+2011MNRAS410.1197}. In the very centre, as we do not detect the Keplerian signature of the SMBH, one could argue that a different $M/L_I$ and thus mass distribution within the hole would obviate the need for a SMBH. To achieve this effect through additional stellar mass would require $M/L_I$ to suddenly increase from 5.7 to $\approx7$. There is however no evidence of a commensurate change in the stellar population \mbox{\citep{Davis+2017MNRAS464.453}}. Additionally, the nuclear activity provides strong evidence for the presence of a SMBH, so we reject this argument.

\subsection{Discussion}
Having considered the sources of systematic uncertainty in Section \ref{sec_sysErrors}, we conclude that the dominant sources of uncertainty are the poorly-constrained inclination and the distance adopted. The remaining uncertainties are the formal errors associated with the model fit. In Section \ref{ssec_modifiedBayes}, we argued that rescaling $\Delta\chi^2$ is required to yield physically reasonable formal uncertainties (and demonstrated that these uncertainties are consistent with those estimated by bootstrapping). We now combine the formal and inclination uncertainties to yield our final result.

We thus seek to combine our best-fitting SMBH mass at fixed-inclination, $M_\mathrm{BH} = 4.0^{+1.6}_{-1.5}\times10^8\,$M$_\odot$, where the uncertainties are $3\sigma$ formal errors, and the uncertainty in the inclination, $20\pm5\degree$. For each model in the final MCMC chain, we draw an inclination from a Gaussian distribution with a $3\sigma$ width of $5\degree$, and then use Equation \ref{eq_MbyInc} to transform the SMBH mass and stellar mass-to-light ratio to the new inclination. The final mass and uncertainty are then the median and $99.7\%$ confidence intervals in the resultant distribution. Our resulting SMBH mass measurement is then \bfMdot, where the uncertainty given is at the $3\sigma$ level.

This mass is approximately half that found by \cite{Krajnovic+2009MNRAS399.1839} using stellar dynamics, but these results are consistent within the $3\sigma$ confidence intervals. The \Msig relation of \cite{McConnellMa2013APJ764.184} yields a SMBH mass estimate of $3.6^{+1.2}_{-1.0}\times10^8\,\Msol$ ($68\%$ confidence interval including intrinsic scatter), compared to which our result is not only consistent, but also very similar. However, the uncertainties on our mass are larger than other CO dynamical measurements. This is due to the combination of the central hole, within which only $\approx22$ per cent of the dynamical mass is contributed by the SMBH, and the poor inclination constraints, the latter a result of the low inclination of the source.

Applying the same prescription to include the effect of inclination on the stellar \mbox{$M/L_I = 5.7\pm0.3$ M$_\odot$/L$_{\odot, I}$} ($3\sigma$ formal uncertainties), we derive that \mbox{$M/L_I = 5.7^{+3.9}_{-1.9}\,$M$_\odot$/L$_{\odot,I}$}. In Section \ref{ssec_MLGrad}, we argued that an increase in the stellar $M/L$ to $M/L_I = 7\,\Msol/L_{\odot,I}$ could obviate the need for a central SMBH, and at face value this inclination-adjusted result is now consistent with no SMBH. However, as we vary the inclination the SMBH mass also increases, so we still robustly recover an SMBH. Our $M/L_I$ is comparable to the dynamical results of \cite{Cappellari+2006MNRAS366.1126} and \cite{Krajnovic+2009MNRAS399.1839} and the stellar population analysis of \cite{Davis+2017MNRAS464.453}. It is however significantly higher than the latter's dynamical mass-to-light ratio, most likely as a result of the different inclination assumed. Using an equation analogous to Equation \ref{eq_MbyInc} to correct this dynamical mass-to-light ratio to the same inclination as the other works we find consistent results.

If our model fully explained our data, we would expect to find the $\chi^2$ value of the best-fitting model to be $\nu\pm\sqrt{2\nu}$, where $\nu$ is the number of degrees of freedom and $\nu\approx N$ due to the large number of data points. We in fact find our best fitting model has a larger $\chi^2$, indicating there are features in the data unexplained by our relatively simple model. This will be common in the exquisite data available in the ALMA era. In the remainder of this paper, we thus will explore features for which our simple model does not account.

\section{Modelling complex molecular gas morphologies with \texttt{SkySampler}}
\label{sec_SSFit}
Our high angular resolution ALMA data reveal an inner hole and much sub-structure in the molecular gas disc of \mbox{NGC524}. In spite of crudely assuming a smooth gas distribution, we were still able to well-constrain the SMBH mass. Previous works on this galaxy used models with one fewer parameter to describe the disc morphology, lacking the truncation radius needed to account for the (then unresolved) central hole.

As ALMA reveals ever more substructures in the discs of many WISDOM targets, such coarse models will not remain appropriate. Yet in some of these, the velocity field still exhibits regular kinematics where gas is present to trace it, implying dynamical modelling is worth pursuing. We therefore face two options - either select arbitrarily complex models, adding free parameters until the gas distributions are well-described, or use the observed gas distributions as inputs to our models, thereby constraining the total flux at each location.

The latter option can be implemented by a simple modification of our existing methods. When we produce the \texttt{KinMS} model of the data cube, we generate a large number of particles with positions $(r,\theta)$ relative to the galaxy centre, such that the density of particles is proportional to the parameterised gas distribution function. We then calculate line-of-sight velocities based on the (axisymmetric) potential specified by our adopted mass model (usually, but not exclusively an MGE model of the stars). However, there is no reason why the gas distribution must be specifiable analytically by a small number of parameters, as we can define an arbitrary set of particle positions that satisfy any density distribution. As long as this (non-axisymmetric) gas distribution is not a significant contributor to the total mass distribution, the assumption of axisymmetry for the potential will remain valid.

The method adopted to build such a model of the gas distribution is as follows. By integrating along the velocity axis of the data cube (i.e. creating a zeroth moment map), we obtain the desired spatial information on the distribution of the gas, independent from the kinematics. We then sample this image with a large number of particles. Were we to assume some particular inclination and position angle, we could now calculate the intrinsic (rather than projected) positions of the particles in the plane of the galaxy's disc. However, we actually include this deprojection step in our MCMC process, since in general we do not want to assume \textit{a priori} a particular orientation of the gas disc. We thus supply these particles to \texttt{KinMS} using the \texttt{inClouds} variable, whereby a particle distribution can be manually specified rather than generated from an analytic radial profile. The particles are thereafter de-projected into the disc plane at each iteration within the MCMC framework (i.e. for each inclination and position angle), before line-of-sight velocities are calculated and the simulated cube generated using the same method as before.

However, the observations consist of the intrinsic gas distribution convolved by the beam, that is oversampled in the data cube. If we were to sample the cube directly, add the velocity field, and then apply the instrumental effects, we would effectively smear the data twice. We therefore sample instead a point source model of the gas distribution obtained from the \texttt{CLEAN} algorithm \citep{Hogbom1974AAS15.417}. This model is already produced for every channel when imaging the data from $uv$ plane visibilities, and we simply add the CLEAN components from all channels here, ignoring the velocity information. Sampling from the de-convolved cube ensures that our particles are placed at positions unbiased by beam smearing, creating a physically reasonable representation of the underlying intrinsic gas distribution. Beam smearing effects are then applied to each channel after the particle velocities are taken into account.

Our cube sampling code is implemented in Python in the software package \texttt{SkySampler}\footnote{\href{https://github.com/Mark-D-Smith/KinMS-skySampler}{https://github.com/Mark-D-Smith/KinMS-skySampler}}. Having integrated along each spaxel, multiple particles are generated within each pixel, where the number generated is proportional to the intensity of the pixel. This allows us to reproduce emission in multiple channels. \texttt{SkySampler} also allows the use of other weighting schemes, so that, for example, the number of particles generated for each spaxel could be proportional to the width of the emission line in that spaxel.

\subsection{Application to \mbox{NGC 524}}

\begin{figure}
        \centering
	\includegraphics[trim={0cm 1cm 1cm 1cm}, width=\columnwidth]{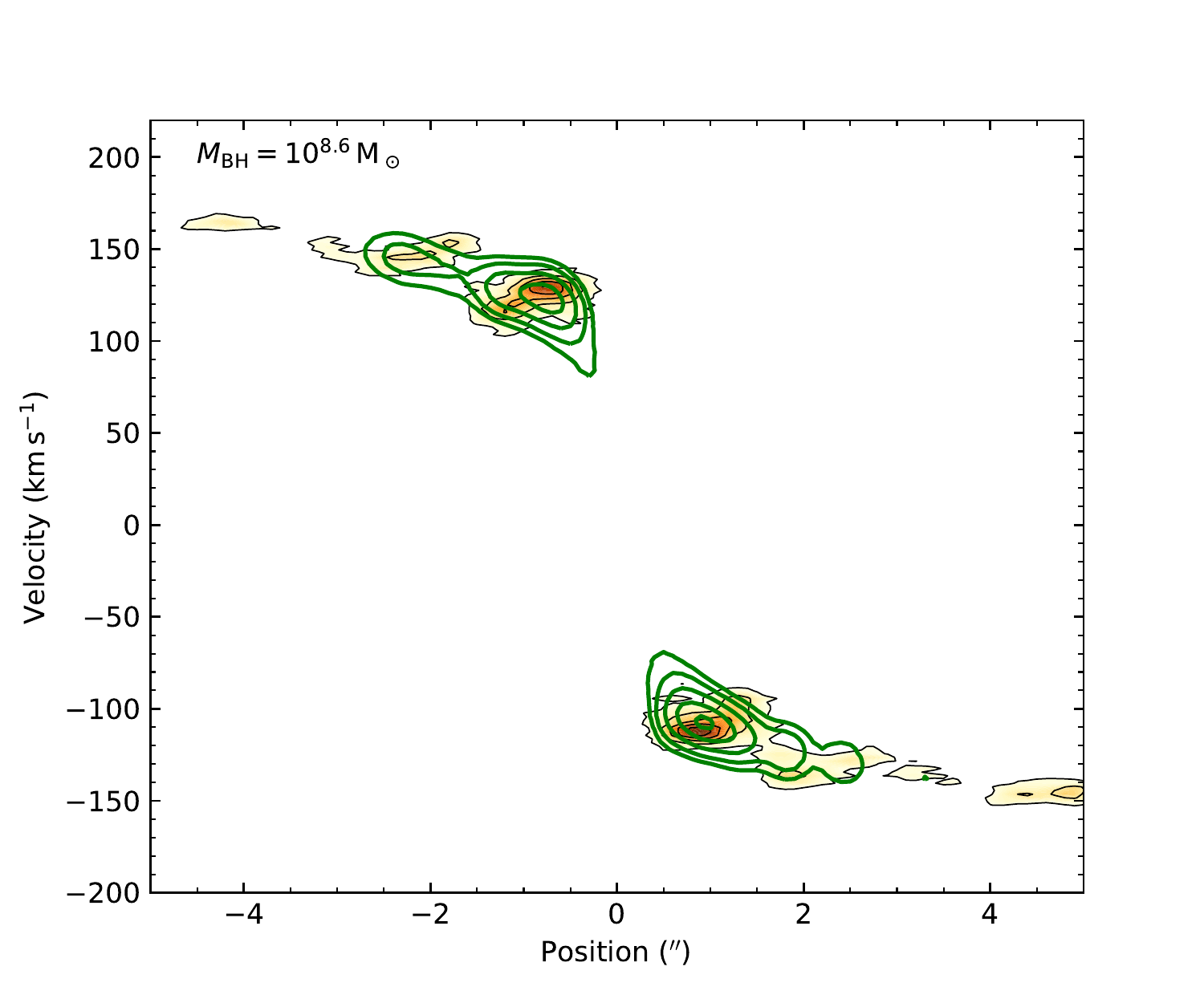}
         \caption{Kinematic major-axis position-velocity diagram for the best-fitting model using a non-parametric description of the gas disc morphology (green contours), overlaid on the observed PVD previously shown in Figure \ref{fig_NGC524_maps} (orange scale and contours). The model only extends to a $3\farcs2$ radius due to the use of the covariance matrix (see Section \ref{ssec_covar}). The SMBH mass and stellar mass-to-light ratio are consistent with those found using a simple parametric model of the gas distribution. The slight deviations of the model from the data at radii $\approx2\arcsec$ are due to the feature described in Section \ref{sec_harmonics}.}
    \label{fig_pvdss}
\end{figure}

Although a good fit can be obtained to the \mbox{NGC 524} data using the simple axisymmetric gas distribution given in Equation \ref{eq_parametricSB}, we can also use the non-parametric distribution generated using SkySampler, allowing us to compare the two methods.

We thus re-run the fixed inclination MCMC fit with the \texttt{SkySampler} model of the gas distribution, that by construction replicates all the rings and holes in the data. Our best-fitting model has $\chi^2_\mathrm{red}=1.82$. The improvement in $\chi^2_\mathrm{red}$ compared to the parametric model (for which $\chi^2_\mathrm{red}=1.84$) might appear modest at first, but it reflects a significant change (roughly 6 times the variance) in the $\chi^2$ value given the large number of pixels in the fit. The posterior is well-constrained, with all parameters consistent with those obtained assuming the parametric gas distribution. The resulting SMBH mass at fixed inclination is $M_\mathrm{BH} = 4.6^{+1.8}_{-1.3} \times 10^8\,$M$_\odot$ ($3\sigma$ formal uncertainties). The associated position-velocity diagram for the best-fitting model is shown in Figure \ref{fig_pvdss} for comparison.

Although for \mbox{NGC 524} the non-parametric method is not strictly required for our fit to converge, this new capability will be useful in sources with complex (non-axisymmetric) gas distributions, enabling us to make measurements without either forcing a coarse parametric model onto the data or requiring an excessive number of free parameters. The efficacy of this method to recover intrinsic source distributions will be tested more formally on simulations in a future paper of this series (North et al.,\,in preparation).

\section{Non-circular motions}
\label{sec_harmonics}
In this section we consider whether our data provide evidence for non-circular motions within the gas, and the effect these might have on the SMBH mass measurement.

\begin{figure}
	\includegraphics[trim={1.5cm 1cm 1.5cm 1cm},width=\columnwidth]{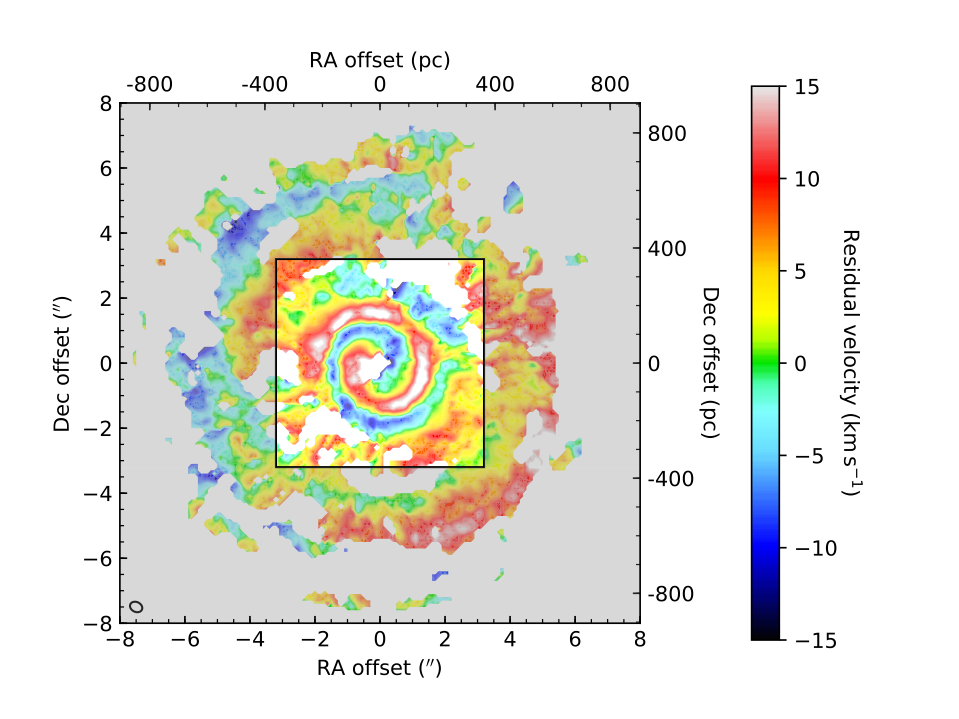}
         \caption{Residuals between the first moments (mean velocity fields) of the data cube and best-fitting model cube. The central box indicates the region within which the model was fit.}
    \label{fig_NGC0524_vresid}
\end{figure}

The model presented in Section \ref{sec_parametricSMBH} assumes all gas is in circular motions, such that the observed velocities are the projection of only an azimuthal component onto the line-of-sight. This appears a good assumption initially, as the observed velocity field shown in the top-left panel of Figure \ref{fig_NGC524_maps} appears smooth, with no significant motion along the minor axis (or equivalently a rather straight zero-velocity curve). However, as previously mentioned, the isovelocity contours at a radius $\approx2\arcsec$ appear distorted, suggesting a purely circular model may be inappropriate. Further evidence for this are revealed by the residual velocity field (Figure \ref{fig_NGC0524_vresid}), generated by subtracting the first velocity moment of a simulated cube (using the best-fitting parameters previously determined, but extending beyond the $6\farcs4\times6\farcs4$ fitting region) from the first velocity moment of the data shown in the top-left panel of Figure \ref{fig_NGC524_maps}. A very clear spiral feature can be seen in the residuals, extending from the central hole with a peak amplitude of $\approx15\,\kms$. The systematic structure of these residuals suggests that the original model is not a complete representation of the data. However, we note that the characteristic magnitude of these residuals is only $\approx10\%$ of the line-of-sight projection of the circular velocity curve, so they probably only trace a small perturbation on top of a dominant axisymmetric potential.

\subsection{Observed velocity field}
We first seek to correlate the distortions of the isovelocity contours and the ring of enhanced velocity dispersions (both visible in Figure \ref{fig_NGC524_maps}) with the residual spiral structure shown in Figure \ref{fig_NGC0524_vresid}. We initially guess that the increased velocity dispersions are due to beam smearing of tightly-spaced isovelocity contours. This broadens the distribution in a single spaxel, as emission in adjacent spaxels at slightly different velocities are blended together. This effect is already seen along the minor axis, where a large gradient in the velocity field as the radius approaches zero causes the isovelocity contours to bunch. Once these are convolved by the beam, emission is smeared over adjacent spaxels, broadening the observed lines.

\begin{figure}
	\includegraphics[trim={0.5cm 2.5cm 0.5cm 2.5cm},width=\columnwidth]{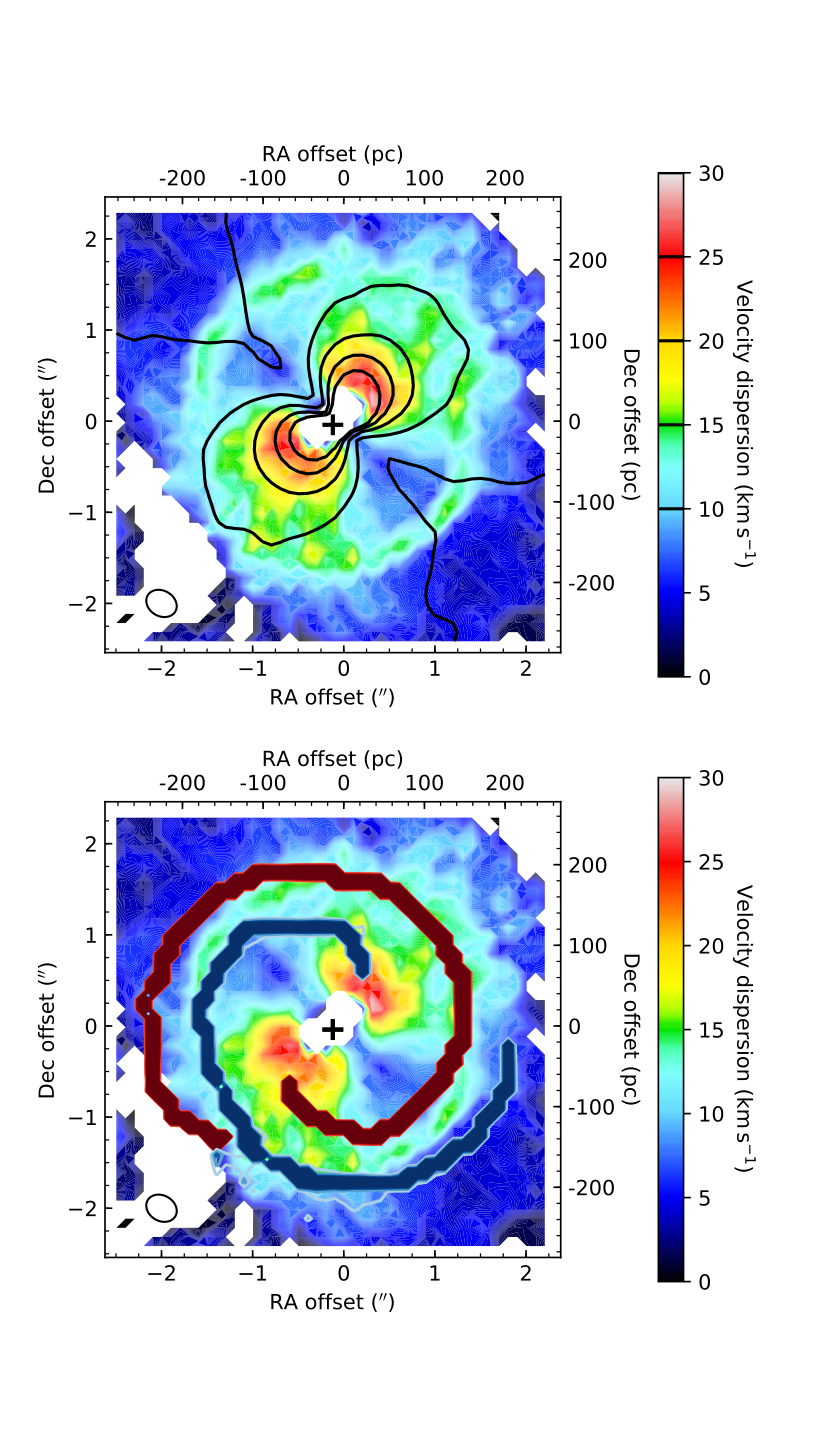}
         \caption{\textbf{Top panel:} Velocity dispersions measured in the best-fitting model data cube (black contours, at $5\,$km$\,$s$^{-1}$ intervals from $10\,$km$\,$s$^{-1}$ overlaid on the observed velocity dispersions (coloured shading). The model cube's velocity dispersions comprise the intrinsic dispersion ($9.3\,$km$\,$s$^{-1}$, as found in Section \ref{ssec_gasVelDisp}) and beam smearing of the circular velocity field. Along the minor axis, the beam smearing accounts well for the observed velocity dispersions, but beam smearing of the model does not account for the excess dispersions along the major axis at $\approx2\arcsec$ radius. \textbf{Bottom panel:}  Red- and blue-shifted arms of the velocity residual shown in Section \ref{fig_NGC0524_vresid} (dark red and dark blue shading and contours), again overlaid on the observed velocity dispersions (coloured shading). The enhanced velocity dispersions on either side of the major axis appear consistent with beam smearing of the two spiral velocity perturbations detected. In both panels, the synthesised beam is shown in the lower-left corner, and the position of the SMBH with a black cross.}
    \label{fig_vDispResid}
\end{figure}

The top panel of Figure \ref{fig_vDispResid} shows the spatial distribution of the observed velocity dispersion (coloured shading), and the predicted velocity dispersion from our best-fitting model cube (black contours), comprising the intrinsic velocity dispersion ($9.3\,$km$\,$s$^{-1}$, as found in Section \ref{ssec_gasVelDisp}) and beam smearing of the circular velocity field. The velocity dispersions in the best-fitting model cube appear to closely match the enhanced velocity dispersion along the minor-axis, validating our assumption that this feature is due to beam smearing of the projected circular velocity field. However, the enhanced velocity dispersions across the major axis at $\approx2\arcsec$ radius are not explained by beam smearing of the projected circular velocity field.

If the increased velocity dispersions at a radius $\approx2\arcsec$ are caused by beam smearing of the two residual arms, the enhanced dispersions will lie between the peaks of the velocity residuals. We initially attempted to parameterise the spiral pattern and fit it with an additional velocity term with a phase described by an Archimedean or logarithmic spiral, but such fits do not match the data. We therefore attempted instead a non-parametric description of the spiral using the ridge-finding algorithm\footnote{\href{https://zenodo.org/record/845874\#.W_2I8Xr7RBw}{DOI:10.5281/zenodo.845874}} of \cite{Steger1998IEEE20.113}, to trace the peaks of the residual image. The residual image was first manually masked to eliminate areas of low signal-to-noise ratios, retaining the principal spiral structure. 

The bottom panel of Figure \ref{fig_vDispResid} again shows the spatial distribution of the observed velocity dispersion (coloured shading), and the velocity residual peaks (red- and blue-shifted arms) identified by the ridge-finding algorithm above. These remaining enhanced velocity dispersions are seen between the two spiral arms, and are on length scales consistent with the synthesised beam. We therefore attribute these enhanced velocity dispersions to beam-smearing of a non-axisymmetric/spiral perturbation in the velocity field.

\subsection{Harmonic expansion}
Having shown that the features identified in Figures \ref{fig_NGC524_maps} and \ref{fig_NGC0524_vresid} are due to a small perturbation on the circular velocity field, we now seek to characterise the nature of this perturbation. In particular, we wish to determine if it is evidence of non-circular motions (i.e. non-zero radial velocities), thereby potentially affecting our SMBH mass measurement. For this, we use the harmonics of the velocity field to separate the azimuthal and radial components of the observed line-of-sight velocities \citep[e.g.][]{Canzian1993ApJ414.487, Schoenmakers+1997MNRAS292.349, Spekkens+2007ApJ664.204}.

We can write the general form of the line-of-sight velocities in terms of the radial and azimuthal components of the gas motion (restricting the motion to the disc plane in the thin disc approximation, and assuming axisymmetry):
\begin{equation}
\label{eq_harmonics}
\frac{v_\mathrm{los}-v_\mathrm{sys}}{\sin i} = \sum_{m=1}^{\infty} c_m(r)\,\cos(m\phi - \phi_0) + s_m(r)\,\sin(m\phi - \phi_0)
\end{equation}
where $\phi - \phi_0$ is the azimuthal phase of a point in the velocity field relative to the position angle $\phi_0$ (the kinematic major axis). The $m=1$ coefficient $c_1$ is then the rotation curve of the galaxy. All higher order terms could, in general, contribute kinematic support to the gas against gravity.

As expected, the phase dependence shows that azimuthal terms are zero along the minor axis, while radial terms are zero along the major axis. A single slit observation of the rotation curve, if not perfectly aligned with the major axis, could yield an incorrect SMBH mass measurement due to artificially reduced velocities, even if no radial component is present. The addition of a radial velocity component will also modify the observed kinematics, affecting the resulting mass model. However, with 3D data and 2D kinematics, as in this work, this problem can be avoided, as the kinematic major axis can be well-determined empirically. The phase difference between the projection of the azimuthal and radial components further allows us to constrain the presence of any inflow/outflow through a harmonic analysis.

\subsubsection{First-order term}
To allow a first-order correction to be included in our models, we re-run the MCMC fit at fixed inclination while allowing an axisymmetric radial component of velocity, that is then projected as an additional contribution to the line-of-sight velocities. This adds one free parameter, the magnitude of this component, that we constrain within priors of $\pm100\,\kms$. Above such a magnitude, this contribution would almost equal the azimuthal component, and therefore be visually obvious in the velocity field. After running the MCMC chains as previously described, the best-fit solution is consistent with no radial flow, and the SMBH mass is unchanged. 

\subsubsection{Higher-order terms}
\label{ssec_harmonicsKinemetry}

\begin{figure}
	\includegraphics[trim={0cm 2.5cm 1.5cm 2cm},width=\columnwidth]{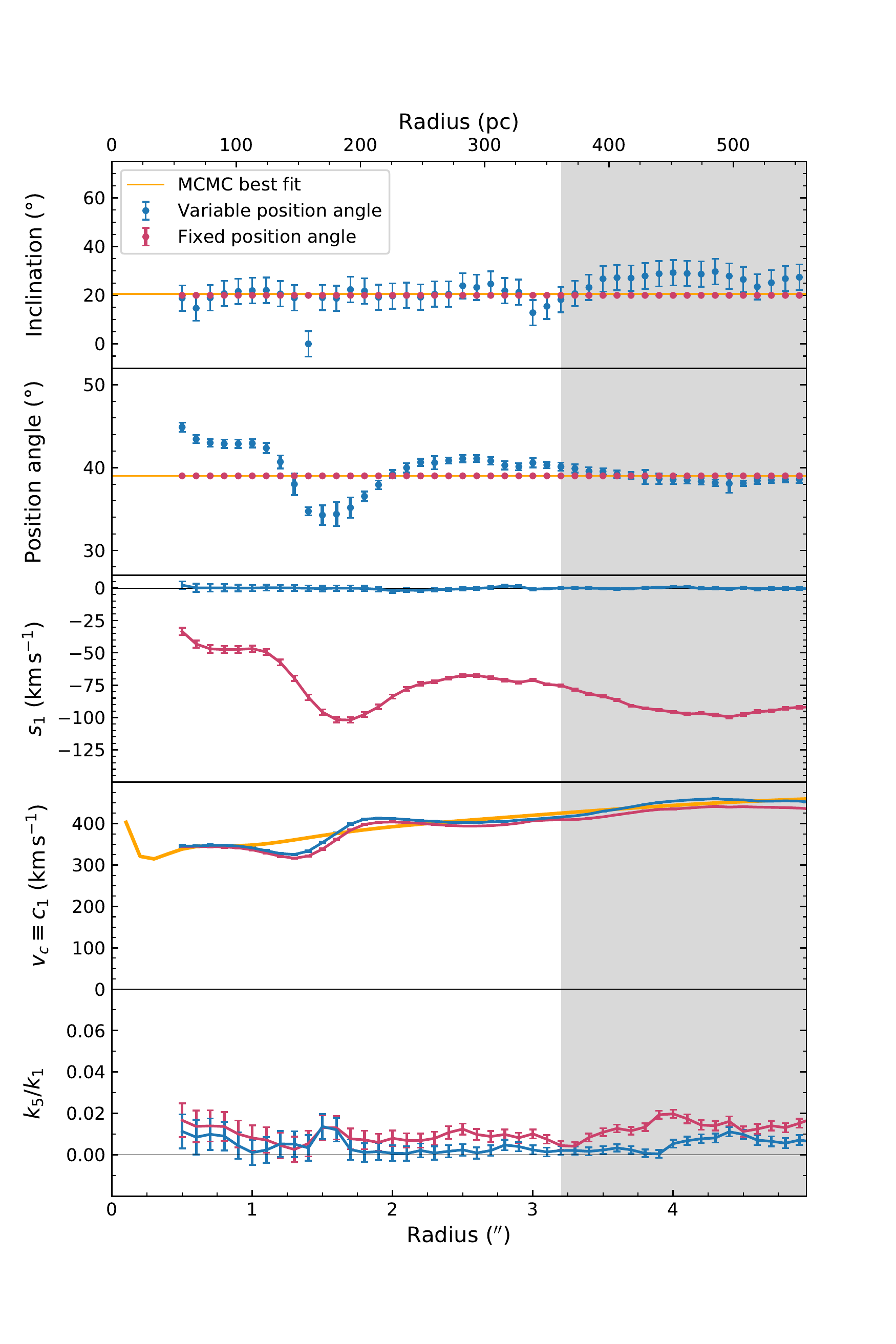}
         \caption{Best-fitting parameters of two harmonic expansions of the observed velocity field, evaluated on ellipses. In blue, the position angles are fit to the velocity field, whereas in red the position angles are fixed to the value used in the full cube fit. The orange line shows the equivalent parameter in the fixed-inclination MCMC fit, while the grey shading shows a radius of $3\farcs2$ that approximates the boundary of the region fit in Section \ref{ssec_SMBHResult}. \textbf{Top panel:} inclinations of the best-fitting ellipses, calculated from Equation \ref{eq_qinc}. \textbf{Second panel:} position angles of the best-fitting ellipses. \textbf{Third and fourth panels:} first-order coefficients $s_1$ and $c_1$, as defined in Equation \ref{eq_harmonics}. \textbf{Lower panel:} higher-order deviations from ordered motion ($k_5\equiv\sqrt{c_5^2+s_5^2.}$).}
    \label{fig_harmonicPA}
\end{figure}

Higher-order harmonic terms in Equation \ref{eq_harmonics} are non-trivial to include in our forward-modelling process, adding many additional parameters to our model. However, higher order harmonics are routinely calculated for 2D velocity fields. Using the \texttt{Kinemetry} package\footnote{\href{http://davor.krajnovic.org/idl/\#kinemetry}{http://davor.krajnovic.org/idl/\#kinemetry}} of \cite{Krajnovic+2006MNRAS366.787} and the observed velocity field (top-right panel of Figure \ref{fig_NGC524_maps}), we fit higher order harmonics to ellipses at the fixed inclination and position angle determined by our MCMC model. The key results are shown in red in Figure \ref{fig_harmonicPA}.

The spiral feature highlighted in Figure \ref{fig_NGC0524_vresid} can also be seen in Figure \ref{fig_harmonicPA} as the $s_1$ term, that from its definition in Equation \ref{eq_harmonics} can be a radial flow. However, the best-fitting ellipse position angles are defined by minimising the $s_1$, $s_3$ and $c_3$ terms of the harmonic expansion \citep{Krajnovic+2006MNRAS366.787}. Thus the spiral feature could also be described as a position angle warp, a possibility not explored in this harmonic expansion nor in the MCMC fits. The degeneracy between a radial flow and a position angle warp in a tilted-ring harmonic expansions of 2D velocity fields was recently also identified and extensively discussed in \cite{Labini+2018arXiv1812.01447}.

To explore the possibility of a position angle warp, we therefore re-run the harmonic expansion in a tilted-ring fit, shown in blue in Figure \ref{fig_harmonicPA}, permitting the model to freely choose both the axial ratio and the position angle of each ring. The inclinations of these rings are calculated from their axial ratios using
\begin{equation}
\cos i \equiv \sqrt{\frac{q^2-q_0^2}{1-q_0^2}}\,\,\,,
\label{eq_qinc}
\end{equation}
where $q$ is the axial ratio of the best-fitting ellipses and $q_0$ is the intrinsic axial ratio of an edge-on galaxy. Since we assume the CO is distributed in a thin disc, the intrinsic axial ratio is $q_0=0$, and Equation \ref{eq_qinc} reduces to $\cos i = q$. The rings have a mean inclination of $21\degree$ and standard deviation of $6\degree$, as previously discussed in Section \ref{ssec_SMBHResult}.

Allowing for a position angle warp leaves no spiral structure after subtracting the first-order harmonics from the observed velocity field. The $k_5 \equiv \sqrt{c_5^2+s_5^2}$ term, the first harmonic that does not determine the parameters of the ellipses, remains small. This suggests that there is no significant radial flow present in the velocity field.

With no significant radial component contributing to the velocity field, we now need to consider what effects the position angle warp may have on the SMBH mass measured. Figure \ref{fig_harmonicPA} shows that the position angle varies radially before settling to the value of $\approx39\degree$ found by the MCMC fit. Centrally, the position angle peaks at $\approx45\degree$. Such a position angle mismatch would mean that the observed central line-of-sight velocities are under-estimates of the true circular velocities, and so we could under-estimate the SMBH mass. Combining Equations \ref{eq_MbyInc} and \ref{eq_harmonics}, we find that the SMBH mass will scale with position angle as:

\begin{equation}
M_\mathrm{BH} \propto v^2 \propto \left(\frac{v_\mathrm{obs}}{\cos{\phi}}\right)^2.
\end{equation}
The position angle warp could thus, at most, increase our SMBH mass by $0.08\,$dex. The actual effect will be smaller than this, since the SMBH mass will be constrained not merely by the central annulus but by all spaxels - the central few providing the strongest constraints. We therefore test the effect of this warp on the SMBH mass measurement by running an additional MCMC chain with the position angle fixed to that found from kinemetry. There is no significant change of the best-fitting SMBH mass, nor of its uncertainties.

\section{Conclusions}
\label{sec_conclusion}
Using high angular resolution ALMA observations of CO(2-1) in the galaxy \mbox{NGC 524}, we have identified a compact $1.3\,\mathrm{mm}$ continuum source, that we find to be spatially consistent with the previously identified compact radio source in this galaxy. Line emission arises from a central molecular gas disc in regular rotation; this disc has a central hole and exhibits a small distortion to the isovelocity contours over the central $2\farcs5$. We showed this distortion can be interpreted as either a position angle warp or evidence for radial flow using a harmonic expansion of the velocity field.

We forward-modelled the kinematics of the gas in the observed cube to measure the SMBH mass. Although the hole prevents us from observing the expected Keplerian increase in the central velocities, we nevertheless obtain a measurement of the supermassive black hole mass of \bfMdot, where the uncertainties stated are at the $3\sigma$ level and include the formal error and the uncertainty in the inclination. The model also yields a stellar mass-to-light ratio of $5.7^{+3.9}_{-1.9}\,\Msol/\mathrm{L_{\odot,I}}$, with uncertainties dominated by the inclination. The formal uncertainties alone in $M/L_I$ are consistent with other results.

The CO disc has a degenerate inclination, so we assumed the gas is coincident with the observed dust, and adopt the inclination previously established from the dust morphology \citep{Cappellari+2006MNRAS366.1126}, that is consistent with a tilted-ring fit to the 2D kinematics from our data. This yielded a good fit to the data, but we subsequently included the effects of this inclination uncertainty in our adopted uncertainties by a Monte Carlo method. The formal uncertainties in our measurement take into account the uncertainty in the $\chi^2$ minimum, and are estimated by bootstrapping and the MCMC process itself, both giving consistent results. The overall uncertainty is dominated by the poorly-constrained inclination.

We also tested whether our result is robust against the gas distribution assumed. The axisymmetric centrally-truncated exponential disc assumed in our original model is only a coarse representation of the underlying morphology, so we introduced a new method to generate a model gas distribution directly from the observations. This model distribution is then kinematically deprojected to model the observed cube. We found no significant change in the best-fitting SMBH mass, but this method will be useful for analysing future observations with complex gas distributions.

Our SMBH mass is consistent with, but half that found using stellar kinematics by \cite{Krajnovic+2009MNRAS399.1839}. It is consistent, and in fact very similar to, that predicted with the \Msig relation of \cite{McConnellMa2013APJ764.184}.

\section*{Acknowledgements}
MDS acknowledges support from a Science and Technology Facilities Council (STFC) DPhil studentship ST/N504233/1. MB was supported by STFC consolidated grant `Astrophysics at Oxford' ST/H002456/1 and ST/K00106X/1. TAD acknowledges support from a STFC Ernest Rutherford Fellowship. MC acknowledges support from a Royal Society University Research Fellowship. KO was supported by Shimadzu Science and Technology Foundation.

This paper makes use of the following ALMA data: ADS/JAO.ALMA\#2015.1.00466.S, ADS/JAO.ALMA\#2016.2.00053.S, ADS/JAO.ALMA\#2017.1.00391.S. ALMA is a partnership of ESO (representing its member states), NSF (USA) and NINS (Japan), together with NRC (Canada), MOST and ASIAA (Taiwan), and KASI (Republic of Korea), in cooperation with the Republic of Chile. The Joint ALMA Observatory is operated by ESO, AUI/NRAO and NAOJ.

This research has made use of the NASA/IPAC Extragalactic Database (NED), which is operated by the Jet Propulsion Laboratory, California Institute of Technology, under contract with the National Aeronautics and Space Administration.

Based on observations made with the NASA/ESA Hubble Space Telescope, and obtained from the Hubble Legacy Archive, which is a collaboration between the Space Telescope Science Institute (STScI/NASA), the Space Telescope European Coordinating Facility (ST-ECF/ESA) and the Canadian Astronomy Data Centre (CADC/NRC/CSA).



\bibliographystyle{mnras}
\bibliography{papers} 


\appendix

\bsp	
\label{lastpage}
\end{document}